\documentclass[12pt]{article}
\usepackage[utf8]{inputenc}
\usepackage{geometry}
\usepackage{graphicx}
\usepackage{array}
\usepackage{subfig}
\usepackage{slashed}
\usepackage{amsmath}
\usepackage{amsfonts}
\usepackage{amssymb}
\usepackage[cm]{fullpage}
\usepackage{cite}
\usepackage{epstopdf}
\usepackage{comment}
\usepackage{hyperref}
\usepackage{cancel}
\usepackage{mathrsfs}

\usepackage{cleveref}
\crefname{section}{§}{§§}
\Crefname{section}{§}{§§}

\usepackage{booktabs} \newcommand{\ra}[1]{\renewcommand{\arraystretch}{#1}}

\numberwithin{equation}{section}

\def\p{\partial}

\def\0{{(0)}}
\def\1{{(1)}}
\def\2{{(2)}}

\def\<{\langle }
\def\>{\rangle }

\newcommand{\bea}{\begin{eqnarray}}

\newcommand{\eea}{\end{eqnarray}}

\newcommand{\be}{\begin{equation}}
\newcommand{\ee}{\end{equation}}
\newcommand{\ba}{\begin{align}}
\newcommand{\ea}{\end{align}}

   \makeatletter
  \let\over=\@@over \let\overwithdelims=\@@overwithdelims
  \let\atop=\@@atop \let\atopwithdelims=\@@atopwithdelims
  \let\above=\@@above \let\abovewithdelims=\@@abovewithdelims
\renewcommand\section{\@startsection {section}{1}{\z@}%
                                   {-3.5ex \@plus -1ex \@minus -.2ex}
                                   {2.3ex \@plus.2ex}%
                                   {\normalfont\large\bfseries}}

\renewcommand\subsection{\@startsection{subsection}{2}{\z@}%
                                     {-3.25ex\@plus -1ex \@minus -.2ex}%
                                     {1.5ex \@plus .2ex}%
                                     {\normalfont\bfseries}}

\newcommand{\beq}{\begin{equation}}
\newcommand{\eeq}{\end{equation}}
\newcommand{\beqa}{\begin{eqnarray}}
\newcommand{\eeqa}{\end{eqnarray}}
\newcommand{\beqar}{\begin{eqnarray*}}

\def\[{\[}
\def\]{\]}

\newcommand{\bd}[1]{\begin{fmffile}{#1}\begin{fmfgraph*}}
\newcommand{\ed}{\end{fmfgraph*}\end{fmffile}}

\begin{document}

\begin{titlepage}

\unitlength = 1mm~\\
\vskip 1cm
\begin{center}

{\LARGE{\textsc{A Solution to the Decompactification Problem\\[0.3cm] in Chiral Heterotic Strings}}}

\vspace{0.8cm}
 {\large Ioannis Florakis}\,{}\footnote{{\tt florakis@lpthe.jussieu.fr}}  {\large and John Rizos}\,{}\footnote{\tt irizos@uoi.gr}

\vspace{1cm}

{\it  ${}^1$ LPTHE, CNRS UMR 7589 - UPMC Paris 6, 4 Place Jussieu, 75005 Paris, France \\
	 ${}^2$ Department of Physics, University of Ioannina, GR45110 Ioannina, Greece 

}

\vspace{0.8cm}

\begin{abstract}
We present a solution to the decompactification problem of gauge thresholds in chiral heterotic string theories with two large extra dimensions, 
where supersymmetry is spontaneously broken by the Scherk--Schwarz mechanism. Whenever the Kaluza--Klein scale that controls supersymmetry breaking is much lower than the string scale, 
the infinite towers of heavy states  contribute non-trivially to the renormalisation of gauge couplings, 
which typically grow linearly with the large volume of the internal space and invalidate perturbation theory. 
We trace the origin of the decompactification problem to properties of the six dimensional theory obtained in the infinite volume limit 
and show that thresholds may instead exhibit logarithmic volume dependence and we provide the conditions for this to occur. 
We illustrate this mechanism with explicit string constructions where the decompactification problem does not occur.

\end{abstract}

\setcounter{footnote}{0}

\vspace{1.0cm}

\end{center}

\end{titlepage}

\pagestyle{empty}
\pagestyle{plain}

\def\vx{{\vec x}}
\def\p{\partial}
\def\po{$\cal P_O$}

\pagenumbering{arabic}

\tableofcontents
\bibliographystyle{utphys}

\section{Introduction}

Understanding the mechanism and implications of supersymmetry breaking in String Theory is an important open problem that may pave the path towards establishing a realistic string phenomenology. Over the years, string constructions with unbroken supersymmetry have been the subject of extensive study, with several seminal results. In the context of string phenomenology, the majority of works in the literature traditionally focused on constructing supersymmetric string vacua with various phenomenologically appealing features, while postponing the actual breaking of supersymmetry to a later stage of the analysis, at the level of the effective field theory. 

Although in some cases this may be sufficient for a tree-level analysis, to actually make contact with low energy data necessarily requires also the proper incorporation of quantum corrections. The latter  receive contributions from the infinite tower of perturbative states of the string and, in practice, are only computable in those cases where an exactly solvable worldsheet CFT description is available. Such is the case of the stringy version \cite{Rohm:1983aq,Kounnas:1988ye,Ferrara:1988jx,Kounnas:1989dk} of the Scherk--Schwarz mechanism \cite{Scherk:1978ta,Scherk:1979zr}, which induces a spontaneous breaking of supersymmetry without spoiling the flatness of the string worldsheet, thanks to the fact that it may be realised as a freely acting orbifold.

The Scherk--Schwarz mechanism corresponds to a flat gauging of supergravity which, in particular, generates a mass term for the gravitino as well as a non-trivial scalar potential of the no-scale type \cite{Cremmer:1983bf}. One of the general features of this mechanism is that the scale of supersymmetry breaking $m_{3/2}\sim 1/R$ is related to the size $R$ of the internal dimensions and remains undetermined at tree level, provided the theory does not encounter tachyonic instabilities. This occurs because the tree-level scalar potential of the theory vanishes upon minimisation with respect to the fields charged under the Scherk--Schwarz gauging of supergravity, leaving the neutral scalars (such as $R$) that enter into the determination of the gravitino mass as free parameters at tree-level.

The situation changes at the loop level, since the scalar potential of the theory receives quantum corrections which typically depend non-trivially on the no-scale moduli. The form of the loop corrected effective potential therefore determines dynamically the size and shape of the internal space and, hence, also the scale of supersymmetry breaking.
Due to the absence of BPS protection, the study of such quantum corrections to the effective potential of string theory with supersymmetry broken by the Scherk--Schwarz mechanism is notoriously non-trivial. This effectively constrains one to work only with asymptotic expressions valid for large volumes of the internal space \cite{Dixon:1990pc}, while requiring an entirely different treatment\footnote{A discussion of issues related to asymptotic volume expansions and how these break down whenever the volume is of the order of the string scale may be found in  \cite{Florakis:2013ura}.} whenever the volume approaches close to the string scale. This is due to convergence issues arising in regions of moduli space close to the string scale, where the contribution of the infinite tower of perturbative string states, including also non-level matched states, become non-negligible. Although several new techniques have been proposed in the recent literature \cite{Angelantonj:2011br,Angelantonj:2012gw,Angelantonj:2013eja,Angelantonj:2015rxa,Florakis:2016boz} that could in principle overcome such issues, they are not yet as efficient for dealing with the effective potential.

Nevertheless, it is possible to infer some of the model independent features of the loop-corrected effective potential by exploiting only the basic properties of the Scherk--Schwarz mechanism. Namely, the potential will necessarily vanish in the infinite volume limit $R\to\infty$, where supersymmetry is recovered \cite{Antoniadis:1990ew}. Secondly, regions close to the string scale $R\sim 1$ (we work in string units $\alpha'=1$) typically include also points fixed under the T-duality group of the theory, which become extrema of the loop-corrected potential \cite{Ginsparg:1986wr}. 

In the simplest cases of supersymmetry breaking via the Scherk--Schwarz mechanism \cite{Itoyama:1986ei,Angelantonj:2006ut}, the potential typically exhibits the form of a puddle, with the minimum at around the string scale $R\sim 1$ and negative values of the potential. This situation is characterised by a number of undesirable features. Firstly, it stabilises the no-scale moduli entering the gravitino mass at the string scale which, in turn, means that supersymmetry is broken at very high energies and therefore, these vacua cannot be used to address the hierarchy problem. Secondly, the value at the minimum is huge and negative compared to the observed value of the cosmological constant, inducing an enormous back-reaction on the classical geometry \cite{Fischler:1986ci,Fischler:1986tb}. Thirdly, whenever the no-scale moduli are stabilised around the string scale, the theory is typically driven into the tachyonic regime, as soon as deformations with respect to the entire set of perturbative moduli of the theory are considered \cite{McClain:1986id,O'Brien:1987pn,Atick:1988si,Kutasov:1990sv,Antoniadis:1991kh,Antoniadis:1999gz,Dienes:1994np,Angelantonj:2008fz,Angelantonj:2010ic,Florakis:2010ty}.

A much more promising scenario \cite{FR}, is one in which the potential instead exhibits the shape of a bump, with its peak centred around the string scale at positive values of the potential, driving the no-scale moduli into the large volume regime $R\gg 1$. In these cases, supersymmetry is naturally broken at scales much smaller than the string scale $m_{3/2}\ll 1$, with a positive value of the cosmological constant and is dynamically protected against the appearance of tachyonic instabilities. 

Additional constraints arise from the asymptotic suppression rate of the potential as $R\gg 1$. For a generic theory in this class, the dominant fall of the one-loop potential is typically polynomial \cite{Antoniadis:1990ew}
\begin{equation}
	V_{\rm eff} \sim \frac{n_F-n_B}{R^4} + \ldots \,,
\end{equation}
where the ellipses denote exponentially suppressed corrections and $n_F, n_B$ are respectively the numbers of massless fermions and bosons in the theory which are unaffected by the Scherk--Schwarz gauging. Assuming the no-scale moduli and, hence, the scale of supersymmetry breaking are eventually stabilised by some non-perturbative mechanism in the TeV range (c.f. the recent discussion in \cite{Abel:2016hgy,Abel:2017rch}), the above naive value of the potential overshoots the observed valued of the cosmological constant by some 34 orders of magnitude.

This leads one to consider models with the additional property of Bose--Fermi degeneracy $n_F=n_F$ in their massless spectra, known in the literature as super no-scale models \cite{Antoniadis:1990ew,Abel:2015oxa,Aaronson:2016kjm,Kounnas:2016gmz,Kounnas:2017mad}. This eliminates the polynomial fall of the potential, which is now forced to fall exponentially fast. The first examples of chiral heterotic models exhibiting this desired behaviour in their potential, together with the additional requirement that the latter has the shape of a bump, were presented in recent work \cite{FR}. Although no comprehensive scan has been performed, the results in \cite{FR} indicate that such models are not rare isolated instances, but actually a plethora of similar super no-scale constructions exists, all sharing the same basic desirable features.

This prompts one to further investigate the properties of such models, focusing in particular on the gauge sector.
In the framework of String Theory, the running of gauge couplings may be significantly affected by the contributions of the massive string modes propagating in the loops. The one-loop corrected gauge coupling $g_a$ associated to a gauge group factor $G_a$ is given at scale $\mu$ by \cite{Kaplunovsky:1987rp}
\begin{equation}
	\frac{16\pi^2}{g_a^2(\mu)} = k_a\,\frac{16\pi^2}{g_s^2} + \beta_a \log\frac{M_s^2}{\mu^2} + \Delta_a \,,
	\label{running}
\end{equation}
where $g_s$ is the  string coupling, $k_a$ is the level of the associated Kac--Moody algebra and $\beta_a$ is the beta function coefficient for the massless states of the theory controlling the familiar logarithmic running involving the ratio of the string scale $M_s$ to $\mu$. As expected, originating from the infrared physics of massless states, the logarithmic contribution is precisely the same as one finds in field theory, with the string scale $M_s$ playing the role of the UV cutoff. The threshold correction $\Delta_a$ encodes the effect of massive string modes, including  oscillators, Kaluza-Klein and winding states. In general, the threshold may be split as
\begin{equation}
	\Delta_a = \tilde\Delta_a(t_i) + \hat\Delta_a \,,
\end{equation}
where $\tilde\Delta(t_i)$ is the contribution arising from subsectors of the theory that effectively enjoy unbroken $\mathcal N= 2$ supersymmetry, or spontaneously broken $\mathcal N=4\to 0$, or $\mathcal N=2\to 0$ supersymmetry, and depends on the compactification moduli $t_i$. The contribution $\hat\Delta$ instead arises from $\mathcal N=1$ subsectors and is necessarily moduli-independent.

This distinction is naturally related to the amount of supersymmetry effectively preserved by the corresponding subsectors, precisely due to the fact that  moduli enter into the threshold corrections through the central charges of the corresponding super-algebras. The moduli independent contribution $\hat \Delta$ is highly model-dependent, but its calculation presents no serious difficulties and is best treated on a case-by-case basis. In what concerns the present work, we shall focus mostly on the model-dependent threshold contributions $\tilde\Delta$ from which the so-called decompactification problem arises \cite{Dixon:1990pc,Antoniadis:1990ew}.

The universal form for differences of gauge thresholds for heterotic strings with spontaneously broken supersymmetry \`a la Scherk--Schwarz was obtained in \cite{Angelantonj:2014dia,Florakis:2015txa,Angelantonj:2015nfa,Angelantonj:2016ibb} and has the form
\begin{equation}
	\begin{split}
	\tilde\Delta_a-\tilde\Delta_b = \sum_i \Bigr\{ -&\alpha_{ab}^i \log\left[ T_2^i U_2^i\, |\eta(T^i)\,\eta(U^i)|^4\right]\\
				-&\beta_{ab}^i \log\left[ T_2^i U_2^i\, |\vartheta_4(T^i)\,\vartheta_2(U^i)|^4\right]\\
				-&\gamma_{ab}^i \log\Bigr[ |\hat j_2(T^i/2)-\hat j_2(U^i)|^4 |j_2(U^i)-24|^4 \Bigr] \Bigr\} \,,
	\end{split}
	\label{thresdiff}
\end{equation}
where $T^i, U^i$ are the K\"ahler and complex structure moduli of the $i$-th $T^2$ torus making up the internal space, $T_2^i={\rm Im}(T^i)$ and $U_2^i={\rm Im}(U^i)$ are the corresponding imaginary parts, $\vartheta_j(\tau)$ are the familiar Jacobi theta constants and $\eta(\tau)$ is the Dedekind function. Furthermore, $j_2(\tau)=(\eta(\tau)/\eta(2\tau))^{24}+24$ is the Hauptmodul of the Hecke congruence subgroup $\Gamma_0(2) \subset {\rm SL}(2;\mathbb Z)$ of the modular group attached to the cusp at infinity, $\hat j_2(\tau) =(\vartheta_2(\tau)/\eta(\tau))^{12}+24$ is its image under the Atkin-Lehner involution $\tau\to-1/(2\tau)$ and, finally, the constants $\alpha_{ab}^i, \beta_{ab}^i$ and $\gamma_{ab}^i$ are recognised as differences of beta function coefficients for the relevant gauge group factors $G_a, G_b$, which encode the entire model dependence of the threshold difference and are determined by the knowledge of the massless spectrum of the theory alone. By exploiting the underlying $\Gamma_0(2)$ modular symmetry, also the gauge thresholds  $\tilde\Delta_{a}$ themselves (as opposed to differences) were  more recently shown in \cite{Florakis:2016aoi}  to admit a universal form in this class of models.

The various contributions in \eqref{thresdiff} can be interpreted as follows. The first line contains the running of couplings in the \emph{exact} $\mathcal N=2$ subsector of the theory. We shall employ term `exact $\mathcal N=2$'  in order to refer to the subsector of the theory with unbroken $\mathcal N=2$ supersymmetry which is  not itself obtained as the (partial) Scherk-Schwarz breaking of an $\mathcal N=4$ theory. Exact $\mathcal N=2$ sectors are central to our subsequent analysis, since they are the ones giving rise to the decompactification problem. In fact, in the infinite volume limit where the Scherk-Schwarz 2-torus decompactifies, one recovers a six dimensional string theory with 8 supercharges, corresponding precisely to the exact $\mathcal N=2$ sector. 

The second line in \eqref{thresdiff} arises from orbifold sectors where supersymmetry is effectively broken as $\mathcal N=4\to 2$. The subsector corresponding to $\mathcal N=4\to 0$ breaking is instead universal and cancels out in threshold differences.  Finally, the third line in \eqref{thresdiff} arises from non-supersymmetric $\mathcal N=2\to 0$ sectors with extra charged massless states under the gauge groups in question, inducing logarithmic singularities at special loci $T^i/2=U^i$ of moduli space, together with their images under the  T-duality group of the theory. It should be stressed that the universal expression \eqref{thresdiff} is quite general and applies also to cases where supersymmetry is left unbroken\footnote{Non-trivial examples with partial spontaneous supersymmetry breaking $\mathcal N=4\to 2$ and non-trivial threshold coefficients $\beta^i_{ab},\gamma^i_{ab}\neq 0$ were presented in \cite{Angelantonj:2015nfa}.}.

Regardless of whether supersymmetry is completely or only partially broken by the Scherk--Schwarz flux, the universal form \eqref{thresdiff} reflects a serious obstacle for realistic string model building with large volume scenarios. Namely, provided the beta function coefficient in the $\mathcal N=2$ subsector of the theory is non-vanishing $\alpha^i_a\neq 0$, and modulo universal\footnote{There are two different notions of universality here. The universality of thresholds or of their differences \eqref{thresdiff}, refers to the functional dependence of thresholds on the compactification moduli $T,U$ and to the fact that model dependence is reduced to a number of constant coefficients. On the other hand, the universal part $Y$ of gauge thresholds refers to the part which does not depend on the gauge group factor and, hence, cancels in differences within a given string model. } contributions $Y$ which are independent of the specific gauge group factor $G_a$, the first line of \eqref{thresdiff} shows that the dominant growth of the threshold is linear as opposed to logarithmic in the volume of the internal tori as $T_2^i\gg 1$
\begin{equation}
	\tilde\Delta_{a} = \alpha_{a}^i\left(\frac{\pi}{3} T_2^i - \log T_2^i\right) +\ldots\, ,
	\label{decompactProb}
\end{equation}
where the ellipses stand for subleading terms. If $\alpha^i_a>0$, the large positive contribution in $\tilde\Delta_a$  forces the running coupling $g_a^2$ in \eqref{running} to become essentially free and $G_a$ then becomes part of the hidden group. In the opposite case, $\alpha^i_a<0$, the huge negative contribution of the threshold drives $g_a^2$ very quickly to the strong coupling regime, unless one adjusts the value of the tree level string coupling $g_s^2$ to unnaturally large precision. Of course, in practice, even this fine-tuning is impossible due to the presence of several such gauge group factors.

This is known as the decompactification problem of gauge couplings and arises whenever the compactification volume is large in units of the string length, $T^i\gg 1$. It is straightforward to see the physical origin of the problem already from \eqref{decompactProb}. When the volume of one of the internal 2-tori becomes large $T_2^i\gg 1$, physics becomes effectively six-dimensional. What \eqref{decompactProb} is effectively doing is  subtracting the usual four-dimensional logarithmic running and replacing it with a six-dimensional one, which has dimensions of length squared.

Whenever at least one supersymmetry is left unbroken by the Scherk--Schwarz mechanism, the scalar potential is protected against radiative corrections and the tree-level compactification moduli $T^i, U^i$ remain moduli to all loop orders. In such situations, the decompactification problem arises only in cases of large volume scenarios. Similarly, in models where supersymmetry is spontaneously broken, but the potential exhibits the form of a puddle, with a minimum at the string scale $T^i \sim 1$, the decompactification problem does not arise, but these models suffer from  Planck scale supersymmetry breaking, backreaction, cosmological constant and tachyon problems as we already mentioned.

Another possibility is to consider models where $\alpha^i_a=0$ such that the linear growth in the volume is absent from the very beginning \cite{Antoniadis:1990ew}. This could be achieved by requiring the absence of the exact $\mathcal N=2$ sector. Namely, by ensuring that any $\mathcal N=2$ subsector in the theory is  obtained as the spontaneous breaking of an $\mathcal N=4$ one, which may be realised by employing freely-acting orbifolds \cite{Kiritsis:1996xd,Faraggi:2014eoa}. In the case of the $\mathbb Z_2^N$-type orbifolds under consideration here, however, this is incompatible with the presence of a chiral matter spectrum, which requires that the original theory (before the Scherk--Schwarz flux is turned on) be realised as an $\mathcal N=1$ preserving non-freely acting orbifold, in which the chiral matter arises precisely from the fixed points. It is then straightforward to show that the $\mathcal N=2$ subsectors of the latter, always give rise to contributions $\alpha^i_a\neq 0$ in \eqref{thresdiff} and, therefore, suffer from the decompactification problem.

The case of interest in this work is the class of chiral models studied in \cite{FR}, which  support the large volume scenario outlined above and for which the decompactification problem becomes very relevant. The purpose of this paper is to show that there exists an interesting, non-trivial class of chiral heterotic theories, with spontaneously broken $\mathcal N=1\to 0$ supersymmetry, with two extra dimensions at naturally large volume that could even be of the TeV range, and with exponentially small positive value for the cosmological constant, for which the decompactification problem does not occur. This is possible thanks to a delicate cancellation in  the volume divergences of  Kaluza--Klein contributions to gauge thresholds  in exact $\mathcal N=2$ sectors, by properly taking into account also the universal contribution $Y$, which has no field theory analogue.

The paper is structured as follows. In Section \ref{SecGrowth}, working in a general setup based on toroidal $\mathbb Z_2$ orbifolds, we identify the sector from which the linear volume growth arises and extract the conditions for it to cancel against the universal gravitational back-reaction $Y$ due to the presence of the gauge fields in Section \ref{SecCondition}. Whenever these conditions are met, thresholds exhibit a logarithmic dependence on the Kaluza-Klein scale and, hence, provide new constraints for realistic string model building. Given the fact that a desert is basically absent in cases when the Kaluza-Klein scale is taken to be in the TeV range, unification occurs directly at the string scale, and in the same section we discuss how such specific string constructions with logarithmic thresholds may be further constrained by known phenomenological parameters at the electroweak scale, such as the Weinberg angle $\sin^2\theta_W$ and the strong coupling constant $\alpha_3(M_Z)$. In Section \ref{SecConstruct}, we trace the origin of the cancellation of linear volume growth in string thresholds to properties of the exact $\mathcal N=2$ subsector, and exploit this fact in order to propose an efficient way to construct $\mathcal N\leq 1$ theories whose gauge couplings do not suffer from the decompactification problem. We end in Section \ref{SecConclude} with our conclusions.

\section{Universality of the dominant growth}\label{SecGrowth}

We begin by extracting the structure of the dominant contributions to gauge thresholds in a model independent way.
For simplicity, our starting point will be the four-dimensional heterotic string theories compactified on $T^6/\mathbb Z_2\times\mathbb Z_2$ orbifolds with $\mathcal N=1$ supersymmetry which may, in particular, accommodate chiral matter. Supersymmetry is then spontaneously broken by the Scherk--Schwarz mechanism, which is itself realised as an additional freely acting $\mathbb Z_2$ orbifold, whose generating element involves the spacetime fermion number parity $(-1)^{F}$, a momentum shift along the first $T^2$ torus of the internal space, and possibly an additional action on the right-moving degrees of freedom ascribed to the Kac-Moody algebra. Furthermore, we shall not be considering continuous Wilson line deformations, since the latter are typically stabilised to vanishing values by the one-loop potential \cite{Kounnas:2016gmz}.

The generic form of the partition function in this class of theories reads
\begin{equation}
		Z = \frac{1}{\eta^{12}\bar\eta^{24}}\,\frac{1}{2^3} \sum_{H_1,h_1,h_2\atop G_1,g_1,g_2}  \hat Z\bigr[^{H_1,h_1,h_2}_{G_1,g_1,g_2}\bigr]\,\Gamma_{2,2}^{(1)}\bigr[^{H_1}_{G_1}\bigr|^{h_1}_{g_1}\bigr](T,U)\,,
	\label{genericPartition}
\end{equation}
where $h_1,h_2$ label the non freely-acting $\mathbb Z_2\times\mathbb Z_2$ orbifold sectors generating the $\mathcal N=1$ theory, $H_1$ labels the orbifold sectors of the freely-acting Scherk--Schwarz orbifold responsible for the breaking of supersymmetry, and the summation over $G_1,g_1,g_2$ enforces the associated orbifold projections. In this expression, we have separated out the contribution of the lattice partition function $\Gamma^{(1)}$ associated to the first $T^2$ torus, which is twisted by the $(h_1,g_1)$ boundary conditions ascribed to the first $\mathbb Z_2$ orbifold, and simultaneously shifted by the Scherk--Schwarz orbifold. It depends on the K\"ahler and complex structure moduli $T$ and $U$ of the first $T^2$ torus, respectively.  All remaining contributions to the partition function are assembled into the block $\hat Z$, which in particular involves the RNS degrees of freedom, the lattice contributions of the remaining two 2-tori which are inert under Scherk--Schwarz, as well as the contribution of the Kac-Moody algebra degrees of freedom.

We note that this generic decomposition encompasses the class of models studied in \cite{FR}, but also includes other possible constructions. For instance, by trivialising one of the non freely-acting orbifolds, e.g. setting $h_1=g_1=0$, the above decomposition describes models where $\mathcal N=2$ supersymmetry is spontaneously broken by the Scherk--Schwarz flux. Similarly, one may even consider supersymmetric vacua by trivialising the Scherk--Schwarz orbifold instead, i.e. $H_1=G_1=0$. Moreover, the generic form of the partition function \eqref{genericPartition} allows the possibility of additional orbifolds, provided their action on the spacetime supercharges and on the Scherk--Schwarz 2-torus is trivial.

The combined action of twists and shifts on the (2,2) lattice may cause some sectors to vanish identically. For example, a lattice that is twisted with respect to a shift orbifold, produces states with non-trivial momentum and winding numbers. If, however, one introduces a simultaneous rotation element of another orbifold inside the trace, the latter will project down to states without momenta or windings, thereby yielding a vanishing contribution. Explicitly, one finds
\begin{equation}
	\Gamma_{2,2}\bigr[^{H}_{G}\bigr|^{h}_{g}\bigr](T,U)= \left\{
				\begin{array}{l l}
							\bigr|\frac{2\eta^3}{\vartheta[^{1-h}_{1-g}]}\bigr|^2 & ,\quad (H,G)=(0,0)\quad,\quad {\rm or}\quad (H,G)=(h,g)\\
							\Gamma^{\rm shift}_{2,2}[^H_G](T,U) & ,\quad h=g=0\\
							0 & ,\quad {\rm otherwise}
				\end{array}
	\right. \,,
	\label{TwiShiLattice}
\end{equation}
and the partition function of the shifted (2,2) Narain lattice with the momentum shift along the first cycle of the $T^2$ is itself defined by
\begin{equation}
	\Gamma^{\rm shift}_{2,2}[^H_G](T,U) = \sum_{m_1,m_2\atop n_1,n_2} (-1)^{G m_1} \ q^{\frac{1}{4}|P_L|^2}\,\bar q^{\frac{1}{4}|P_R|^2} \,,
\end{equation}
with $P_L$, $P_R$ being the complexified lattice momenta
\begin{equation}
	\begin{split}
	P_L = \frac{ m_2 -U m_1 +T(n_1+\frac{H}{2}+U n_2)}{\sqrt{ T_2 U_2}} \,,
	P_R = \frac{ m_2 -U m_1 +\bar T(n_1+\frac{H}{2}+U n_2)}{\sqrt{ T_2 U_2}} \,.
	\end{split}
\end{equation}

Without loss of generality, we may express the one-loop threshold correction $\Delta_a$ in a form analogous to \eqref{genericPartition}
\begin{equation}
	\Delta_a = \int_{\mathcal F} \frac{d^2\tau}{\tau_2}\, \sum_{H_1,h_1,h_2\atop G_1,g_1,g_2}   \Psi_a\bigr[^{H_1,h_1,h_2}_{G_1,g_1,g_2}\bigr]\,\Gamma_{2,2}^{(1)}\bigr[^{H_1}_{G_1}\bigr|^{h_1}_{g_1}\bigr](T,U)\,.
\end{equation}
The integral is performed over the moduli space of the worldsheet torus, with $\mathcal F$ being the canonical fundamental domain $\mathcal F=\mathbb H^+/{\rm SL}(2;\mathbb Z)$ defined as the quotient of the Poincar\'e upper half-plane $\mathbb H^+$ modded out by the modular group ${\rm SL}(2;\mathbb Z)$. We have again separated out the contribution of the shifted/twisted (2,2) lattice associated to the 2-torus on which the Scherk--Schwarz orbifold acts. All other remaining contributions including, in particular, the helicity supertrace as well as the group trace in the gauge sector, have been absorbed into $\Psi_a=\Psi_a(\tau,\bar\tau)$, which is in general a non-holomorphic function of the complex structure $\tau=\tau_1+i\tau_2$ on the worldsheet torus.

In accordance with the analysis in \cite{FR}, we shall assume that only the volume of the Scherk--Schwarz 2-torus may dynamically roll to large values $T_2\gg 1$ and is, therefore, relevant for the decompactification problem. Furthermore, by inspection of eq. \eqref{TwiShiLattice}, it is easy to see that only the sector $h_1=g_1=0$ may lead to contributions to the gauge thresholds that depend on the Scherk--Schwarz moduli. We, therefore, focus on the $T,U$-dependent contribution to $\Delta_a$ 
\begin{equation}
	\tilde\Delta_a = \int_{\mathcal F} \frac{d^2\tau}{\tau_2} \sum_{H_1, G_1}\Gamma_{2,2}^{\rm shift}\bigr[^{H_1}_{G_1}\bigr](T,U)\,  \tilde\Psi_a\bigr[^{H_1}_{G_1}\bigr] \,,
\end{equation}
where $\tilde\Psi_a$ is the combination
\begin{equation}
	\tilde\Psi_a\bigr[^{H_1}_{G_1}\bigr] = \sum_{h_2,g_2}\Psi_a\bigr[^{H_1\,,\,0\,,\,h_2}_{G_1\,,\,0\,,\,g_2}\bigr] \,.
\end{equation}
It is now convenient to organise the sum over $H_1,G_1$ into two ${\rm SL}(2;\mathbb Z)$ orbits. The first one arises from the term $H_1=G_1=0$, which is already invariant under the full modular group and, therefore, constitutes an orbit in itself. The remaining three terms $(H_1,G_1)\neq (0,0)$  close into a separate ${\rm SL}(2;\mathbb Z)$ orbit, which is generated by $(H_1,G_1)=(0,1)$ under the action of the elements $S$ and $ST$ of ${\rm SL}(2;\mathbb Z)$, where $S$ is the inversion $\tau\to-1/\tau$ and $T$ is the unit translation $\tau\to\tau+1$.
Explicitly, this decomposition reads
\begin{equation}
	\tilde\Delta_a = \tilde\Delta_a^{\rm I} + \tilde\Delta_a^{\rm II} \,,
\end{equation}
with the contributions of the two orbits being given by
\begin{equation}
		\tilde\Delta_a^{\rm I} = \int_{\mathcal F} \frac{d^2\tau}{\tau_2}\, \Gamma_{2,2}(T,U)\,  \tilde\Psi_a\bigr[^{\,0\,}_{\,0\,}\bigr] \,,
		\label{IntegralI}
\end{equation}
where we now denote the standard (unshifted) Narain lattice as $\Gamma_{2,2}(T,U)$, and
\begin{equation}
		\tilde\Delta_a^{\rm II} = \int_{\mathcal F_0(2)} \frac{d^2\tau}{\tau_2}\, \Gamma_{2,2}^{\rm shift}\bigr[^{\,0\,}_{\,1\,}\bigr](T,U)\,  \tilde\Psi_a\bigr[^{\,0\,}_{\,1\,}\bigr] \,.
		\label{IntegralII}
\end{equation}
To obtain the above expression for the second orbit $\Delta_a^{\rm II}$, we have changed variables into $\tau'=S\cdot \tau$ and $\tau'=ST\cdot \tau$ in the corresponding terms, and further exploited the modular properties of the integrand in order to undo these transformations. We therefore obtain an integral involving only the generating term $(H_1,G_1)=(0,1)$ in this orbit, at the cost of enlarging the fundamental domain to its image under the $S$ and $ST$ elements of ${\rm SL}(2;\mathbb Z)$,
\begin{equation}
	\mathcal F_0(2) = \mathcal F\cup (S\cdot \mathcal F)\cup (ST\cdot \mathcal F)\,.
\end{equation}
Of course, the integrand of $\Delta_a^{\rm II}$ is no longer invariant under ${\rm SL}(2;\mathbb Z)$ but only under its Hecke congruence subgroup $\Gamma_0(2)$.

In order to study the large volume limit $T_2\gg 1$, it is most convenient to Poisson resum the lattice momenta $m_1,m_2\in\mathbb Z$ and cast the (2,2) lattice into its Lagrangian representation. For the lattices relevant for $\tilde\Delta_a$, this reads
\begin{equation}
	\Gamma_{2,2}^{\rm shift}\bigr[^{\,\,\,0\,}_{\,G_1\,}\bigr] = \frac{T_2}{\tau_2}\sum_{m_i,n_i\in\mathbb Z} e^{-2\pi i T\,\det A-\frac{\pi T_2}{\tau_2 U_2}\left|(1,U)A\binom{\tau}{1}\right|^2} \,,
	\label{LatticeLagrange}
\end{equation}
where $A$ is the winding matrix 
\begin{equation}
	A= \begin{pmatrix}
					n_1 & m_1+\frac{1}{2}G_1 \\
					n_2 & m_2\\
	\end{pmatrix} \,.
\end{equation}
The presence of the lattice shift in orbit II ($G_1=1$) is responsible for the two orbits exhibiting entirely different behaviour in their threshold contributions at large volume. This is most easily seen at the level of the lattice itself. In the large volume limit, the $n_i$ winding modes are suppressed since they become highly massive and therefore are taken to vanish $n_i=0$, while the Kaluza--Klein states tend to accumulate and produce a quasi-continuous spectrum. In the case of orbit I, all entries of the winding matrix $A$ are integral, which implies that the dominant term in the unshifted lattice comes from the configuration $A=0$, and the lattice grows linearly in $T_2$. Different is the case of orbit II, where the half-unit momentum shift along the first cycle now implies that the winding matrix $A$ is always non-vanishing and, therefore, the shifted lattice vanishes exponentially as $T_2\gg 1$.

Although this simple argument was based on the properties of the Narain lattice \eqref{LatticeLagrange}, the same basic conclusions can be drawn by explicitly evaluating the modular integrals \eqref{IntegralI} and \eqref{IntegralII} via lattice unfolding and carefully extracting their large volume behaviour. One then finds that a possible linear growth in $T_2$, which would give rise to the decompactification problem, can only arise from the sector $H_1=G_1=0$ corresponding to the first orbit $\tilde\Delta_a^{\rm I}$, while $\tilde\Delta_a^{\rm II}$ instead exhibits at most logarithmic growth. 

We shall, therefore, focus our attention on the first orbit, $H_1=G_1=0$, which does not feel the spontaneous breaking of supersymmetry of the Scherk--Schwarz orbifold. As we mentioned already, in order for the contribution $\tilde\Delta_a^{\rm I}$ to be non-trivial, this sector must enjoy exact $\mathcal N=2$ supersymmetry. Clearly, if it would also preserve $\mathcal N=4$ supersymmetry, the corresponding contributions to gauge thresholds would vanish and, hence, the non-trivial case of interest here is from  $(h_2,g_2)\neq(0,0)$, involving 
\begin{equation}
	\tilde\Psi_a\bigr[^{\,0\,}_{\,0\,}\bigr] = \sum_{(h_2,g_2)\neq (0,0)}\Psi_a\bigr[^{0\,,\,0\,,\,h_2}_{0\,,\,0\,,\,g_2}\bigr] \,.
\end{equation}
By construction, the block $\tilde\Psi_a\bigr[^{\,0\,}_{\,0\,}\bigr]$ entering \eqref{IntegralI} is invariant under the full modular group ${\rm SL}(2;\mathbb Z)$. Given that this sector preserves an exact $\mathcal N=2$ supersymmetry, one may immediately see that it also possesses special quasi-holomorphy properties. Indeed, this sector is BPS saturated and its contributions are related to the modified elliptic genus \cite{Harvey:1995fq,Harvey:1996gc}. This property forces the left-moving conformal weights of BPS states to be determined entirely by the $\mathcal N=2$ central charge, which is identified with the left-moving lattice momenta, $m_{\rm BPS}^2=|P_L|^2$. The left-moving string oscillators are then constrained to provide exactly half a unit of conformal weight and, modulo the (2,2) lattice contribution, the entire $q$-dependence of the integrand is washed out. Technically, this is the result of a cancellation of the holomorphic RNS contributions in the helicity supertrace and of additional oscillator terms against the holomorphic part of the (4,4) lattice twisted by the $(h_2,g_2)$ orbifold. 

Even though $\tilde\Psi_a\bigr[^{\,0\,}_{\,0\,}\bigr]$ does not involve powers of $q=e^{2\pi i \tau}$ in its Fourier expansion, it is still not a holomorphic object. This is because loop corrections to gauge couplings receive also universal contributions which essentially do not depend on the choice of gauge group, but which break holomorphy by introducing factors of $1/\tau_2$. From the point of view of a background field method calculation, these arise from  gravitational back-reaction to the background gauge fields. From the point of view of a scattering amplitude calculation, they arise as contact terms, from points on the worldsheet torus where  vertex operators corresponding to the external states collide. Technically, the quasi-holomorphy enters into the group traces from the current-current correlator
\begin{equation}
	\int_\Sigma d^2 z\,\langle J^a(z) J^a(0)\rangle  =  {\rm Tr}\,Q_a^2-\frac{k_a}{4\pi\tau_2} \,,
	\label{contactterm}
\end{equation}
where $Q_a^2$ is the charge and $k_a$ the level of the corresponding Kac-Moody algebra. Importantly, this implies that at most a single power of $1/\tau_2$ may appear in the gauge thresholds $\Delta_a$.

Modularity and quasi-holomorphy severely constrain the form of the $\tilde\Psi_a\bigr[^{\,0\,}_{\,0\,}\bigr]$ and, in fact, dictate that it be an element of the polynomial ring of weak, quasi-holomorphic, modular invariant functions. As a result, it admits a unique expansion into the generators of the ring,
\begin{equation}
	\tilde\Psi_a\bigr[^{\,0\,}_{\,0\,}\bigr] =  k_a A\,\frac{\hat{\bar E}_2 \bar E_4 \bar E_6}{\bar \Delta}+B_a\, \frac{\bar E_4^3}{\bar\Delta}+C_a \,,
	\label{ExpandPsi}
\end{equation}
where $A,B_a,C_a$ are constants, $E_4$ and $E_6$ are the weight 4 and 6 holomorphic Eisenstein series, respectively, $\Delta(\tau)=\eta^{24}(\tau)$ is the weight 12 cusp form (known as the modular discriminant), and $\hat E_2$ is the weight 2 quasi-holomorphic Eisenstein series. The latter is the modular completion of the holomorphic, but quasi-modular, weight 2 Eisenstein series, $\hat E_2 = E_2-3/(\pi\tau_2)$. Note that the holomorphic Eisenstein series we consider here are normalised canonically, i.e. $E_w(\tau)=1+\mathcal O(q)$. In view of \eqref{contactterm}, the term involving $\hat E_2$ depends on the gauge group factor only by the multiplicative level factor $k_a$, while $A$ is a universal factor, independent of the choice of gauge group which, a priori, may be model-dependent.

Following \cite{Kiritsis:1996dn}, the above structure \eqref{ExpandPsi} for the $\mathcal N=2$ contribution of interest, based on holomorphy and modularity, implies that the only possible singularity around $\bar q=0$ is of the form of a simple pole. This state corresponds precisely to the ubiquitous unphysical tachyon of the heterotic string (also known as proto-graviton). However, this state is neutral with respect to all gauge group factors and should, therefore, introduces no singularity at special loci in the bulk of the $(T,U)$ moduli space. The absence of the simple pole in $\bar q$ determines $B_a=- k_a A$. Furthermore, the constant term in \eqref{ExpandPsi} is identified as the beta function coefficient $\hat\beta_a$ for the exact $\mathcal N=2$ sector, which gives $C_a=\hat\beta_a+1008 k_a A$. Putting everything together, \eqref{ExpandPsi} becomes
\begin{equation}
	\tilde\Psi_a\bigr[^{\,0\,}_{\,0\,}\bigr] =  k_a A\left(\frac{\hat{\bar E}_2 \bar E_4 \bar E_6-\bar E_4^3}{\bar \Delta}+1008\right)+\hat\beta_a \,.
\end{equation}

The universal coefficient $A$ may now be determined by repeating the same analysis for gravitational thresholds \cite{Kiritsis:1996dn,Antoniadis:1992sa,Antoniadis:1992rq,Kiritsis:1994ta,Florakis:2016aoi} and identifying the analogous contribution from the exact same $\mathcal N=2$ sector under consideration here, giving rise to
\begin{equation}
	\tilde\Psi_{\rm Grav}\bigr[^{\,0\,}_{\,0\,}\bigr] = A\, \frac{\hat{\bar E}_2 \bar E_4 \bar E_6}{\bar \Delta} \,,
\end{equation}
involving the same coefficient $A$. The constant term in the $\bar q$-expansion of $\tilde\Psi_{\rm Grav}$ is identified as the conformal anomaly of the corresponding $\mathcal N=2$ subsector, $\hat\beta_{\rm Grav}=-264A$ and may be easily extracted from the massless spectrum \cite{Antoniadis:1992sa,Antoniadis:1992rq}. Computing it for a simple ${\rm K3}\times T^2$ compactification, with the K3 surface realised in its singular limit as a $T^4/\mathbb Z_2$ orbifold, one finds $\beta_{\rm Grav}^{{\rm K3}\times T^2}=22$. This differs from our case by a factor of 4, i.e. $\hat\beta_{\rm Grav}=\beta_{\rm Grav}^{{\rm K3}\times T^2}/4$ due to the presence of two additional $\mathbb Z_2$ orbifold factors: the orbifold $(h_1,g_1)$ reducing supersymmetry down to $\mathcal N=1$ and the Scherk--Schwarz orbifold $(H_1,G_1)$ responsible for the spontaneous breaking of supersymmetry down to $\mathcal N=0$. Taking this into account, the value of the universal constant is found to be $A=-1/48$ for all models in this class.

Separating the universal contribution proportional to the level $k_a$ from the running of the couplings, proportional to $\hat\beta_a$, we may write \cite{Kiritsis:1996dn}
\begin{equation}
	\tilde\Delta_a^{\rm I} = -\frac{k_a}{48}\, Y + \hat\beta_a\, \Delta \,,
	\label{thresholdDecomposition}
\end{equation}
where the universal part $Y$ is defined as \cite{Petropoulos:1996rr,Henningson:1996jz}
\begin{equation}
	Y = \int_{\mathcal F}\frac{d^2\tau}{\tau_2}\,\Gamma_{2,2}(T,U)\left(\frac{\hat{\bar E}_2 \bar E_4 \bar E_6-\bar E_4^3}{\bar \Delta}+1008\right) \,,
	\label{Yuniversal}
\end{equation}
while the integral associated to the running may be easily computed in closed form \cite{Dixon:1990pc} in terms of Dedekind functions
\begin{equation}
	\Delta = {\rm R.N.}\int_{\mathcal F}\frac{d^2\tau}{\tau_2}\,\Gamma_{2,2}(T,U) = - \log\left[4\pi e^{-\gamma}\, T_2 U_2\,\left|\eta(T)\,\eta(U)\right|^4\right]\,,
	\label{DKL}
\end{equation}
where $\gamma$ is the Euler-Mascheroni constant.
Naturally, the latter integral is IR divergent due the presence of massless states and should be defined by means of an appropriate renormalisation scheme (R.N.). In this section, it is convenient to employ the renormalisation scheme introduced in \cite{Angelantonj:2012gw,Angelantonj:2011br}. The final result differs from the one traditionally used in the string literature by additive numerical constants. For convenience, we shall restore the constants relevant to the $\overline{DR}$ scheme in the following section, although they are irrelevant for the discussion of the decompactification problem itself.

The universal part $Y$ may also be explicitly computed (c.f. \cite{Angelantonj:2012gw,Angelantonj:2015rxa}) and, working in the Weyl chamber of interest $T_2>U_2$, one finds
\begin{equation}
	Y = 6 \log\left|j(T)-j(U)\right|^4 + \frac{16\pi}{T_2}\,E(2;U) + 2\sum_{M>0}\frac{q_T^M}{M}\left(1+\frac{1}{\pi M T_2}\right)\,T_M\cdot \mathcal F(2,1,0;U)+{\rm c.c.}\,,
	\label{Yeval}
\end{equation}
where $j(\tau)=E_4^3/\Delta$ is the Klein invariant $j$-function, $E(s;\tau)$ is the non-holomorphic Eisenstein series of ${\rm SL}(2;\mathbb Z)$,
\begin{equation}
	E(s;U) = \frac{1}{2}\sum_{m,n\in\mathbb Z\atop (m,n)\neq(0,0)} \frac{\tau_2^s}{|m+\tau n|^{2s}}\,,
\end{equation}
the nome $q_T$ associated to the $T$ modulus is defined as $q_T=e^{2\pi i T}$, and $T_M$ is the $M$-th Hecke operator acting on the Niebur-Poincar\'e series $\mathcal F(2,1,0;U)$ as \cite{Angelantonj:2012gw}
\begin{equation}
	T_M\cdot \mathcal F(2,1,0;U) = \sum_{a,d>0\atop ad=M}\sum_{b\,{\rm mod}\, d}\mathcal F\left(2,1,0;\frac{aU+b}{d}\right) \,.
\end{equation}
Finally, the function $\mathcal F(2,1,0;U)$ is identified \cite{Angelantonj:2012gw} as the weak, quasi-holomorphic modular function 
\begin{equation}
	\mathcal F(2,1,0;\tau) = \frac{ \hat E_2 E_4 E_6+5E_4^3}{\Delta}+264 \,. 
\end{equation}
Alternative expressions in terms of polylogarithms may also be obtained \cite{Kiritsis:1997hf,Angelantonj:2015rxa}, but their explicit form is less compact than \eqref{Yeval}.

We shall not dwell deeper into the structure and properties of $Y$ as given in \eqref{Yeval}, since we are only interested in its large volume limit. A direct inspection  reveals that only the first term in eq.  \eqref{Yeval}, involving the logarithm of the difference of $j$-functions, grows in the large volume limit. The second term exhibits instead power like suppression, while the last two terms involving the sum of non-vanishing frequencies $M\neq 0$ are always exponentially suppressed as $T_2\gg 1$. Thus, keeping only the linear growth, we find
\begin{equation}
	Y = 48\pi T_2 + \mathcal O(T_2^{-1}) \,.
\end{equation}
Similarly, we may extract the growth of $\Delta$ in \eqref{DKL} and find
\begin{equation}
	\Delta = \frac{\pi}{3}\,T_2 -\log T_2 +\mathcal O(e^{-2\pi T_2}) \,.
\end{equation}
Plugging these into \eqref{thresholdDecomposition}, we may finally assemble the leading behaviour of gauge thresholds for all models in this class in the large volume limit 
\begin{equation}
	\Delta_a = \left(\frac{\hat \beta_a}{3}-k_a\right)\pi T_2 + \mathcal O(\log T_2) \,.
	\label{ThresLinearGrowth}
\end{equation}
An alternative way to extract the linear growth of the thresholds is to replace the unshifted (2,2) lattice $\Gamma_{2,2}(T,U)$ in \eqref{Yuniversal} and \eqref{DKL} with the volume factor $T_2$, arising from the term where the matrix of windings vanishes, $A=0$, and then use Stokes' theorem, or the techniques of \cite{Angelantonj:2012gw} to evaluate
\begin{equation}
	\int_{\mathcal F}\frac{d^2\tau}{\tau_2^2}\left(\frac{\hat{\bar E}_2 \bar E_4 \bar E_6-\bar E_4^3}{\bar \Delta}+1008\right) = 48\pi \,,
\end{equation} 
in the case of $Y$, while for $\Delta$ the analogous integral involves simply the volume of the ${\rm SL}(2;\mathbb Z)$ fundamental domain, equal to $\pi/3$.

Although our primary motivation is to apply the above considerations to the models examined in \cite{FR}, where the Kac-Moody algebras are constructed at level $k_a=1$, the above analysis is also relevant for more general constructions where some of the gauge group factors may be realised at higher level.

\section{Theories with logarithmic thresholds}\label{SecCondition}

The form \eqref{ThresLinearGrowth} suggests a possible way out of the decompactification problem. Indeed, one may imagine a situation where the  beta function $\hat\beta_a$ of the exact $\mathcal N=2$ subsector precisely cancels against the universal part,
\begin{equation}
	\hat\beta_a =3k_a \,,
	\label{Condition}
\end{equation}
for all relevant gauge group factors $G_a$ of the theory including, in particular, those in the observable sector. Whenever the decompactification equality \eqref{Condition} is saturated, the corresponding couplings will exhibit logarithmic volume dependence that is determined by the remaining sectors in the theory, including also the contribution of the twisted matter originating from the $\mathcal N=1$ sectors. Of course, gauge group factors in the hidden sector may still violate the decompactification condition \eqref{Condition}, as long as their $\mathcal N=2$ beta functions dominate over the universal part, $\hat\beta_a > 3k_a$, in which case their couplings are essentially driven to the free regime.

The fact that the decompactification condition \eqref{Condition} imposes $\hat\beta_a$ to be a positive number by no means implies that the associated gauge group is non-asymptotically free. We stress here that $\hat\beta_a$ is the beta function only in the exact $\mathcal N=2$ subsector defined by $H_1=G_1=h_1=g_1=0$ and $(h_2,g_2)\neq(0,0)$. Condition \eqref{Condition} then ensures that the volume divergence in the running $\Delta$ of the gauge couplings is cancelled by the back-reaction term $Y$. What remains is a logarithmic running of the gauge couplings with logarithmic thresholds which, in the $\overline{DR}$ scheme takes the form
\begin{equation}
	\frac{16\pi^2}{g_a^2(\mu)}= k_a\frac{16\pi^2}{g_s^2} + \beta_a\, \log\frac{M_s^2}{\mu^2} + \beta_a' \,\log\left(\frac{2e^{1-\gamma}}{3\pi\sqrt{3}}\frac{M_{\rm KK}^2}{M_s^2}\right) + \ldots
	\label{logrunning}
\end{equation}
controlled by effective beta functions $\beta_a, \beta_a'$ that are determined from the massless states in the remaining sectors. 
Here,  $\beta_a$ is the beta function associated to the massless states in the theory from all sectors, whereas $\beta_a'$ is the beta function contribution of those massless states for which a Kaluza-Klein tower from the Scherk--Schwarz 2-torus is allowed. Furthermore,  $M_{KK}=1/\sqrt{T_2}$ is the Kaluza-Klein scale, while the ellipses denote both terms that are suppressed as $\sim 1/T_2$ as well as exponentially suppressed contributions. Clearly, for an asymptotically free gauge group factor, $\beta_a<0$.

The origin of the coefficients $\beta_a, \beta'_a$ may be further clarified by explicitly identifying the orbifold sectors from which they arise. To this end, we define the following constant terms in the $q,\bar q$-series
\begin{equation}
	\begin{split}
	&b_a^{(1)} = \sum_{(h_2,g_2)\neq(0,0)}\Psi_a\bigr[^{0,0,h_2}_{0,0,g_2}\bigr]\Bigr|_{\rm cst} \,,\\
	&b_a^{(2)} = \sum_{h_2,g_2}\Psi_a\bigr[^{0,0,h_2}_{1,0,g_2}\bigr]\Bigr|_{\rm cst} \,,\\
	&b_a^{(3)} = \sum_{(h_1,g_1)\neq(0,0)}\sum_{h_2,g_2}\left(\Psi_a\bigr[^{0,h_1,h_2}_{0,g_1,g_2}\bigr] +\Psi_a\bigr[^{h_1,h_1,h_2}_{g_1,g_1,g_2}\bigr]\right)\Bigr|_{\rm cst} \,,\\
	\end{split}
	\label{b123}
\end{equation}
The first line denotes the contribution from the exact $\mathcal N=2$ sector that leaves the Scherk--Schwarz lattice unaffected, and which generically gives rise to the decompactification problem, $b_a^{(1)}\equiv\hat\beta_a$.  The term $h_2=g_2=0$ in the second line corresponds to the spontaneous breaking $\mathcal N=4\to 0$ of the theory, while the terms $(h_2,g_2)\neq(0,0)$ correspond to subsectors $\mathcal N=4\to 2$ (for $h_2=0$ and $g_2=1$) and $\mathcal N=2\to 0$ (otherwise). For both $b_a^{(1)}$ and $b_a^{(2)}$, the Scherk--Schwarz lattice is untwisted (and shifted for $b_a^{(2)}$) and involves the contribution from the Kaluza-Klein towers of states. 

Different is the case of the third line in \eqref{b123}, which corresponds to the cases where the Scherk--Schwarz lattice is twisted, and which includes also the $\mathcal N=1$ sectors. The first term inside the sum, for which $H_1=G_1=0$, is essentially the supersymmetric $\mathcal N=1$ theory before the Scherk--Schwarz breaking takes place. The second term in the summand corresponds to the subsector in which the Scherk--Schwarz orbifold effectively merges together with the $(h_1,g_1)$ one, and loses its freely-acting property. We can then identify the beta function coefficients entering \eqref{logrunning} as the combinations
\begin{equation}
	\beta_a = b_a^{(1)}+b_a^{(2)}+b_a^{(3)}\quad,\quad \beta_a' = b_a^{(1)}+b_a^{(2)}\,.
\end{equation}
Of course, given that the form \eqref{logrunning} already assumes that the decompactification condition \eqref{Condition} is met for the gauge group factors of interest, we have $b_a^{(1)}=3k_a$. 

Before discussing any specific construction, it is instructive to see some of the generic consequences of string models that do not suffer from the decompactification problem. To this end, we shall assume that some realistic string model has been constructed, such that  \eqref{Condition} is valid for the Standard Model gauge group factors, while the hidden sector ones satisfy instead $\hat\beta_a > 3k_a$ and will be ignored for the purposes of this analysis. 
 
We define, as usual, $\alpha_a^2=g_a^2/4\pi$ for $a=Y,2,3$, to be the coupling parameters of the hypercharge U(1), SU(2) and  SU(3) gauge groups, respectively.   We may now obtain a useful relation for the string coupling $\alpha_s^2=g_s^2/4\pi$ which does not involve the electroweak values of the Standard Model couplings $\alpha_i(M_Z)$ 
\begin{equation}
	\frac{k_2+k_Y}{\alpha_s} = \frac{1}{\alpha_{\rm em}}-\frac{\beta_2+\beta_Y}{4\pi}\,\log\frac{M_s^2}{M_Z^2}-\frac{\beta_2'+\beta_Y'}{4\pi}\log\left(\frac{2e^{1-\gamma}}{3\pi\sqrt{3}}\,\frac{M_{\rm KK}^2}{M_s^2}\right) \,,
\end{equation}
but instead expresses it in terms of the Kaluza-Klein scale, the electromagnetic ${\rm U}(1)_{\rm em}$ coupling constant $\alpha_{\rm em}(M_Z)=e^2/4\pi$ and  the beta functions $\beta,\beta'$ of the theory.
Similarly,  we obtain the Weinberg angle
\begin{equation}
	\sin^2\theta_W= \frac{k_2}{k_2+k_Y}+ \frac{\alpha_{\rm em}}{4\pi}\left[ \frac{k_Y\beta_2-k_2\beta_Y}{k_2+kY}\,\log\frac{M_s^2}{M_Z^2}+ \frac{k_Y\beta_2'-k_2\beta_Y'}{k_2+kY}\,\log\left(\frac{2e^{1-\gamma}}{3\pi\sqrt{3}}\,\frac{M_{\rm KK}^2}{M_s^2}\right) \right] \,,
\end{equation}
and the SU(3) coupling constant at the electroweak scale
\begin{equation}
	\begin{split}
	\frac{1}{\alpha_3(M_Z)}&=\frac{k_3}{\alpha_{\rm em}(k_2+k_Y)}\\
					&+\frac{1}{4\pi}\left[\left(\beta_3-\frac{k_3(\beta_2+\beta_Y)}{k_2+k_Y}\right)\log\frac{M_s^2}{M_Z^2}+\left(\beta_3'-\frac{k_3(\beta_2'+\beta_Y')}{k_2+k_Y}\right)\log\left(\frac{2e^{1-\gamma}}{3\pi\sqrt{3}}\frac{M_{\rm KK}^2}{M_s^2}\right)\right] \,.
	\end{split}
\end{equation}
From the fact that the Einstein-Hilbert term does not renormalise at one-loop in the case of Scherk--Schwarz supersymmetry breaking \cite{Kiritsis:1994ta,Florakis:2016aoi}, the value of the string coupling may be further related to the Planck scale.

\begin{figure}
\centering
\includegraphics[width=0.6\textwidth]{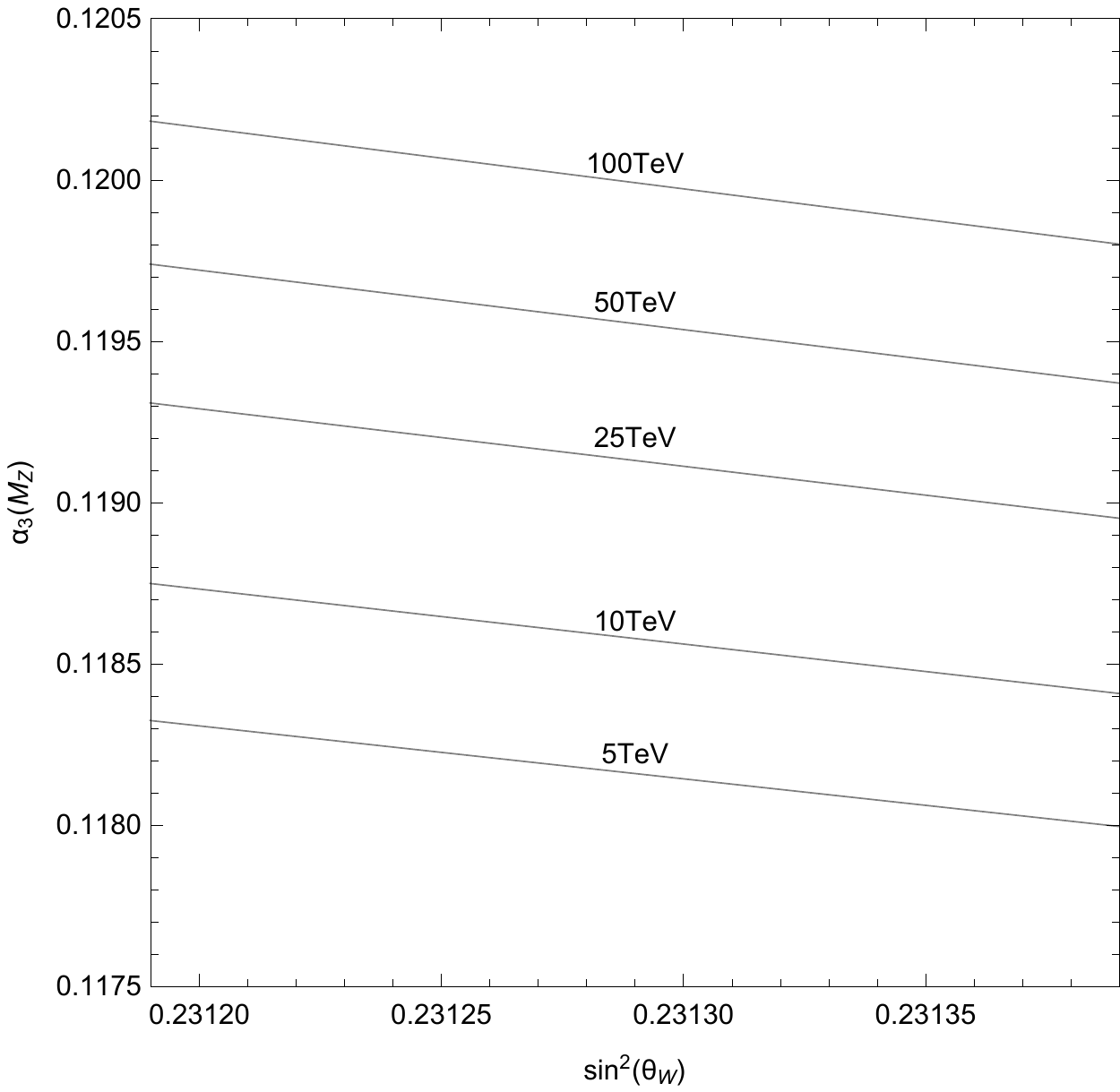}
\caption{\label{fig}{\it Contour plot of the Kaluza--Klein scale $M_{\rm KK}$ against the predicted values of the $\rm SU(3)$ coupling and the Weinberg angle. The string coupling $a_s$ ranges between $0.083$ (for $\rm M_{KK}=100$ TeV) and $0.108$ (for $\rm M_{KK}=5$ TeV).}}
\end{figure}

As an illustrative example, let us consider an idealised minimal scenario in which the full string construction is such that $(\beta_Y,\beta_2,\beta_3)=(-7,-\frac{19}{6},\frac{41}{6})$ are set equal to the corresponding values in the Standard Model, $(k_Y,k_2,k_3)=(\frac{5}{3},1,1)$ and  $(\beta'_Y,\beta_2',\beta_3')=(-\frac{15}{2},-\frac{43}{6},-\frac{23}{3})$. It is then possible to show that there exist regions in the Kaluza-Klein parameter space, consistent with the string unification picture\footnote{Here, the term string unification refers to the universal string coupling at tree level.}, which are compatible with experimental bounds for both $\sin^2\theta_W$ and $\alpha_3(M_Z)$, as depicted in Figure \ref{fig}.


\section{Explicit constructions}\label{SecConstruct}

So far, we have discussed in some generality the sector which potentially causes string thresholds to grow linearly with the volume of the Scherk--Schwarz torus and pointed out the existence of a class of theories in which the decompactification problem does not occur. In this section, we wish to illustrate this approach with some explicit examples. For the purposes of our discussion, we will concentrate on the class of  models analysed in \cite{FR}. 

The specific example we shall consider here is further characterised by super no-scale structure, which guarantees an exponentially small value for the cosmological constant.  In fact, its potential is identical\footnote{The model we present in this section is not identical to Model C in \cite{FR}, even though their partition functions and, therefore, their one-loop corrected scalar potentials are indeed numerically equal. However, such models typically differ in both their spectra and interactions.} to that of the meta-stable ``Model C'' presented in \cite{FR} and, therefore, dynamically leads to the large volume scenario discussed in the Introduction, in which the decompactification problem becomes \emph{a priori} particularly relevant. 
The model itself is a four-dimensional heterotic theory with spontaneously broken $\mathcal N=1\to 0$ supersymmetry, obtained as a $T^6/(\mathbb Z_2)^6$ orbifold  with net matter chirality $N=4$. The compactification then breaks the original ${\rm E}_8\times{\rm E}_8$  gauge group down to ${\rm SO}(10)\times {\rm U}(1)^3\times {\rm SO}(8)^2$.

The action of the six $\mathbb Z_2$ orbifold factors explicitly reads
\begin{equation}
	\begin{split}
			&\mathbb Z_2^{(1)} \ :\ (-1)^{F_2}\,r_1 \ ,\ r_1\,:\,\{ X^{1,2,5,6}\to-X^{1,2,5,6}\}+{\rm standard\,\, embedding}\,,\\
			&\mathbb Z_2^{(2)} \ :\ (-1)^{F_2}\,r_2 \ ,\ r_2\,:\,\{ X^{3,4,5,6}\to-X^{3,4,5,6}\}+{\rm standard\,\, embedding}\,,\\
			&\mathbb Z_2^{(3)} \ :\ (-1)^{F_{\rm s.t.}+F_1+F_2}\,\delta_1 \ ,\ \delta_1\,:\,\{ X^1\to X^1+\pi R_1\}\,,\\
			&\mathbb Z_2^{(4)} \ :\  \delta_2:\,\{ X^3\to X^3+\pi R_3\}\,,\\
			&\mathbb Z_2^{(5)} \ :\ (-1)^{F_1+F_2}\,\delta_3 \ ,\ \delta_3\,:\,\{ X^5\to X^5+\pi R_5\}\,,\\
			&\mathbb Z_2^{(6)} \ :\ (-1)^{F_1}\,r_3 \ ,\ r_3\,:\,\{ \bar\phi^{5,6,7,8}\to-\bar\phi^{5,6,7,8}\} \,,\\
	\end{split}
	\label{OrbAction}
\end{equation}
where $X^i$ are the worldsheet coordinates spanning the internal $T^6$ space, $R_i$ are the radii of the corresponding cycles, and $\bar\phi^{1},\bar\phi^{2},\ldots, \bar\phi^{8}$ are the right-moving worldsheet scalars generating the level 1 Kac-Moody algebra of the hidden sector gauge group ${\rm SO}(8)\times {\rm SO}(8)'$, in the construction of  \cite{FR}. The spacetime fermion number is denoted by $F_{\rm s.t.}$, while $F_1, F_2$ are the corresponding `fermion numbers' ascribed to the two ${\rm E}_8$ factors of the original ${\rm E}_8\times{\rm E}_8$ heterotic string in ten dimensions, before compactification.

The action of the first two $\mathbb Z_2$ factors is not free, and these precisely construct the unbroken $\mathcal N=1$ theory, in which  chiral matter arises from the fixed points. The third $\mathbb Z_2$ factor is identified as the freely acting Scherk--Schwarz orbifold, responsible for the spontaneous breaking of supersymmetry $\mathcal N=1\to 0$, inducing the non-trivial gravitino mass $m_{3/2}=|U|/\sqrt{T_2 U_2}$, where $T,U$ are the K\"ahler and complex structure moduli of the first 2-torus along which the momentum shift $\delta_1$ is acting. The fourth and fifth $\mathbb Z_2$ factors are similarly freely acting, whereas the sixth $\mathbb Z_2$ is non-freely acting and is responsible for splitting the hidden sector ${\rm E}_8$ into ${\rm SO}(8)\times{\rm SO}(8)'$. In addition to the above action, the model also involves the following choice of discrete torsion
\begin{equation}
	\epsilon(1,2),\,\epsilon(1,4),\,\epsilon(2,3),\,\epsilon(2,4),\,\epsilon(2,6),\,\epsilon(3,4),\,\epsilon(4,6),\,\epsilon(5,6)\,.
\end{equation}

The one-loop partition function of the model at the generic point in the perturbative heterotic moduli space reads
\begin{align}
	\begin{split}
	Z = &\frac{1}{\eta^{12}\bar\eta^{24}}\,\frac{1}{2^3}\sum_{h_1,h_2,H\atop g_1,g_2,G}\frac{1}{2^3}\sum_{a,k,\rho\atop b,\ell,\sigma} \frac{1}{2^3}\sum_{H_1,H_2,H_3\atop G_1,G_2,G_3} (-1)^{a+b+HG+\Phi} \vartheta[^a_b]\,\vartheta[^{a+h_1}_{b+g_1}]\,\vartheta[^{a+h_2}_{b+g_2}]\,\vartheta[^{a-h_1-h_2}_{b-g_1-g_2}] \\
			&  \times  \Gamma^{(1)}_{2,2}[^{H_1}_{G_1}|^{h_1}_{g_1}](T^{(1)},U^{(1)})\ \Gamma^{(2)}_{2,2}[^{H_2}_{G_2}|^{h_2}_{g_2}](T^{(2)},U^{(2)})\ \Gamma^{(3)}_{2,2}[^{H_3}_{G_3}|^{h_1+h_2}_{g_1+g_2}](T^{(3)},U^{(3)}) \\
		& \times	\ \bar\vartheta[^k_\ell]^5 \,\bar\vartheta[^{k+h_1}_{\ell+g_1}]\,\bar\vartheta[^{k+h_2}_{\ell+g_2}]\,\bar\vartheta[^{k-h_1-h_2}_{\ell-g_1-g_2}]
			 \,\bar\vartheta[^{\rho}_{\sigma}]^4 \,\bar\vartheta[^{\rho+H}_{\sigma+G}]^4 \,.
	\end{split}
	\label{orbifoldpartitionf}
\end{align}
Here, $a,b$ are the spin structures associated to the RNS worldsheet fermions of the orbifold theory, with the NS sector corresponding to $a=0$ and the R sector to $a=1$. 
The 16 complex fermions realising the level one Kac-Moody algebra of each ${\rm E}_8$ factor of the ten-dimensional heterotic string carry $\mathbb Z_2$-valued boundary conditions $(k,\ell)$ and $(\rho, \sigma)$.

Similarly, $(h_1, g_1)$ and $(h_2,g_2)$ are the parameters for the $\mathbb Z_2^{(1)}\times\mathbb Z_2^{(2)}$ non-freely acting orbifolds with standard embedding generating a chiral $\mathcal N=1$ theory which may be thought of as the singular limit of a Calabi-Yau manifold. Namely, the  $h_i=0,1$ label the various orbifold (un)twisted sectors, while summing over $g_i$ imposes the corresponding  $\mathbb Z_2$ projections. Furthermore, $H_i, G_i=0,1$ are generically associated to the three freely-acting  orbifold factors $\mathbb{Z}_2^{(3)},\mathbb Z_2^{(4)},\mathbb Z_2^{(5)}$, each involving an order-two momentum  shift on the corresponding $T^2$ torus. The Scherk--Schwarz breaking $\mathbb Z_2^{(3)}$ is triggered by the momentum shift along the a-cycle of the first 2-torus and, hence, its $T^{(1)},U^{(1)}$ moduli become the relevant parameters for the decompactification problem. As in \cite{FR}, we will henceforth adopt the simplified notation $T^{(1)}\equiv T$ and $U^{(1)}\equiv U$, while keeping the moduli of the remaining tori fixed at the fermionic point. Finally, $H,G=0,1$ are associated to the $\mathbb Z_2^{(6)}$ orbifold twisting 4 Kac-Moody currents in the hidden ${\rm E}_8$ directions and breaking it down to ${\rm SO}(8)\times {\rm SO}(8)'$.

A specific choice for the modular invariant phase $\Phi$ uniquely fixes the model in question, by properly implementing the action  \eqref{OrbAction} of the $T^6/\mathbb Z_2^6$ orbifold and implements, in particular,  the Scherk--Schwarz breaking of supersymmetry through the correlation of the spacetime R-symmetry charges to the momentum and winding charges of the first 2-torus. In our particular example, it reads 
\begin{align}
	\begin{split}
	\Phi &= ab + k\ell + \rho \sigma\\
		   & + (aG_1+b H_1+H_1 G_1)+(ag_2+b h_2+h_2 g_2)\\
		   & + (kG+\ell H+HG)+( kg_1+\ell h_1+h_1 g_1) +( kG_1+\ell H_1+H_1 G_1) \\
		   & +( kg_2+\ell h_2+h_2 g_2)+( kG_3+\ell H_3+H_3 G_3)\\
		   & + (\rho g_1+\sigma h_1+h_1 g_1)+( \rho G_1+\sigma H_1+ H_1 G_1)\\
		   & + (\rho g_2 +\sigma h_2+h_2 g_2)+ (\rho G_3+ \sigma H_3+H_3 G_3)\\
		   & + (g_2 H + h_2 G)+(G_2 H+H_2 G)+(G_3 H+H_3 G)\\
		   & + (g_2 h_1+ h_2 g_1)+(g_2 H_1+h_2 G_1)+(G_2 H_1+H_2 G_1)\\
		   & + (G_2 h_2+H_2 g_2)+ (g_1 H_2+ h_1 G_2) \,.
	\end{split}
\end{align}

We are now ready to analyse the gauge thresholds in this model. It is certainly possible to perform an exhaustive analysis, based on the technique of decomposition into modular orbits \cite{Angelantonj:2013eja,Florakis:2016aoi} generated by elements of congruence subgroups of ${\rm SL}(2;\mathbb Z)$, and obtain the full threshold corrections in the form of series expansions valid for large volume $T_2\gg 1$. However, for the sake of simplicity, we choose to extract only the dominant contribution of the thresholds at large volume.

To this end, we first define 
\begin{equation}
	\begin{split}
	B_a[^{H_1}_{G_1}|^{h_1}_{g_1}]&= \frac{U[^{h_1}_{g_1}]}{2^9\pi} \sum_{a,k,\rho\atop b,\ell,\sigma}\sum_{\gamma_2,\gamma_3 \atop \delta_2,\delta_3}\sum_{h_2,H\atop g_2, G} (-1)^{a+b+HG+\Phi}\, \frac{\Lambda[^a_b]}{\eta^2}\,\frac{\vartheta[^a_b]\vartheta[^{a+h_1}_{b+g_1}]\vartheta[^{a+h_2}_{b+g_2}]\vartheta[^{a-h_1-h_2}_{b-g_1-g_2}]}{\eta^4}   \\
	\times&  \frac{|\vartheta[^{\gamma_2}_{\delta_2}]\vartheta[^{\gamma_3}_{\delta_3}]\vartheta[^{\gamma_2+h_2}_{\delta_2+g_2}]\vartheta[^{\gamma_3+h_1+h_2}_{\delta_3+g_1+g_2}]|^2}{\eta^6 \bar\eta^6} \,
	\mathcal D_a\left[ \frac{\bar\vartheta[^k_\ell]^5 \bar\vartheta[^{k+h_1}_{\ell+g_1}] \bar\vartheta[^{k-h_1}_{\ell-g_1}] \bar\vartheta[^{k-h_1-h_2}_{\ell-g_1-g_2}] \bar\vartheta[^{\rho}_{\sigma}]^4 \bar\vartheta[^{\rho+H}_{\sigma+G}]^4 }{\bar\eta^{18}}\right]\Bigr|_{z=0} 
	\end{split}
	\label{FullThresholdEx}
\end{equation}
where
\begin{equation}
	U[^h_g] = \left\{  \begin{array}{c l}
					1 & ,\ {\rm if} \ h=g=0\\
					\left|\frac{2\eta^3}{\vartheta[^{1-h}_{1-g}]}\right|^2 & ,\ {\rm if}\ (h,g)\neq(0,0)
				\end{array} \right. \,,
\end{equation}
and $\Lambda$  effectively inserts the helicity charge in the supertrace
\begin{equation}
	\Lambda[^a_b]=\partial \log\left(\frac{\vartheta[^a_b]}{\eta}\right)\,.
\end{equation}
The group trace is realised in terms of the differential operator $\mathcal D_a =\frac{1}{(2\pi i)^2}\,\partial_{z_a}^2$
acting on the Jacobi parameter $z_a$ of the theta function $\bar\vartheta[^{\alpha}_{\beta}](z|\tau)$ associated to the Cartan direction of the group whose thresholds we are interested in. Namely, in writing \eqref{FullThresholdEx}, it is implied that one of the right-moving theta constants $\bar\vartheta[^{\alpha}_{\beta}]\equiv \bar\vartheta[^{\alpha}_{\beta}](0|\bar\tau)$ inside the brackets, is in fact replaced by the corresponding theta function $\bar\vartheta[^{\alpha}_{\beta}](z|\bar\tau)$, such that the action of $\mathcal D_a$ precisely inserts the square of the Cartan charge inside the group trace. For example, in the case of SO(10), the relevant replacement is $\bar\vartheta[^k_\ell]^5 \to \bar\vartheta[^k_\ell]^4 \bar\vartheta[^k_\ell](z|\bar\tau)$.

Using \eqref{b123}, the beta function coefficients of the model can then be straightforwardly computed by extracting the constant terms
\begin{equation}
	(b_a^{(1)}, b_a^{(2)}, b_a^{(3)} )= \biggr( B_a[^0_0|^0_0] \quad,\quad B_a[^0_1|^0_0] \quad,\quad \sum_{(h_1,g_1)\neq(0,0)}( B_a[^0_0|^{h_1}_{g_1}]+B_a[^{h_1}_{g_1}|^{h_1}_{g_1}]) \biggr)\biggr|_{q,\bar q=0} \,.
\end{equation}
A straightforward calculation then yields 
\begin{equation}
	\begin{split}
	{\rm SO}(10) \quad &: \quad (b_{10}^{(1)},b_{10}^{(2)}, b_{10}^{(3)}) = (3, -\tfrac{31}{3},-4) \\
	{\rm SO}(8) \quad &: \quad (b_{8}^{(1)},b_{8}^{(2)}, b_{8}^{(3)}) = (3, -\tfrac{23}{3},4) \\
	{\rm SO}(8)' \quad &: \quad (b_{8'}^{(1)},b_{8'}^{(2)}, b_{8'}^{(3)}) = (3, -5,-4) \\
	\end{split}
\end{equation}
As is the case with all models in the class studied in \cite{FR}, all non-abelian gauge group factors are realised at Kac-Moody level $k_a=1$ and, therefore, the present model satisfies the decompactification condition \eqref{Condition}, i.e. $b_a^{(1)}=3$. Although our specific example is simply a toy model, since it is based on the unbroken SO(10) observable group, it still serves to illustrate the salient features of our mechanism. 

Following the discussion of the previous section, we hence see that the decompactification problem does not occur in this model and the thresholds behave logarithmically as in  \eqref{logrunning}, with the beta function coefficients being given by
\begin{equation}
	\begin{split}
	{\rm SO}(10) \quad &: \quad (\beta_{10}, \beta'_{10}) = (-\tfrac{34}{3} , -\tfrac{22}{3})\\
	{\rm SO}(8) \quad &: \quad (\beta_{8},\beta'_{8}) = (-\tfrac{2}{3}, -\tfrac{14}{3})\\ 
	{\rm SO}(8)' \quad &: \quad (\beta_{8'},\beta'_{8'}) = (-6,-2) 
	\end{split}
\end{equation}
Even though the absence of the decompactification problem requires $b_a^{(1)}=3k_a$ to be positive, this does not imply that the associated gauge group factor is non asymptotically-free. Indeed, as mentioned in previous sections, the decompactification condition \eqref{Condition} only originates from the very specific exact $\mathcal N=2$ subsector of the theory. Whether a gauge group factor is asymptotically free or not is dictated by $\beta_a$, which receives contributions from all sectors of the theory, including the $\mathcal N=1$ sectors where chiral matter arises. This is the case in our particular toy model, where all three non-abelian gauge groups have negative beta functions.

The particular model we have constructed does indeed have the desired property of eliminating the linear growth from its thresholds and, therefore, does not suffer from the decompactification problem. We wish to stress that this is not an accidental property of this specific model. In fact, it is possible to construct several different theories with   spontaneously broken or even unbroken supersymmetry, without ever encountering the decompactification problem.  A preliminary computer scan over  the class of models discussed in \cite{FR}, comprising $10^{11}$ vacua, and for the subclass (of approximately $7\times 10^4$ models)  that meet a minimal set of phenomenological criteria, including the absence of tachyons and the presence of chiral fermions, yields the results of Table \ref{mytable} for the $SO(10)\times{SO(8)}\times{SO(8)'}$
group factors. As can be seen from the first line of Table \ref{mytable}, around 40\% of ``phenomenologically" acceptable models in this class do not
suffer from the decompactification problem, at least for the three non-abelian gauge couplings considered here. What is more, there exist examples in which one may  arrange for nearly all $U(1)$ factors to also satisfy \eqref{Condition}.  Since the linear volume growth originates only from subsectors with exact $\mathcal N=2$ supersymmetry, it is clear that the decompactification problem will be absent in any theory for which the corresponding $\mathcal N=2$ subsector satisfies $\hat\beta_a=3k_a$.

\begin{table*}\centering
\ra{1.3}
\begin{tabular}{@{}rrr|rr@{}}\toprule
$\hat{\beta}_{10}$&$\hat{\beta}_{8}$&$\hat{\beta}_{8'}$&\# of models&\%\\ \midrule
3&3&3&29456&38.5\\ 
9&$-3$&$-3$&15840&20.7\\ 
$-3$&9&9&14000&18.3\\ 
\bottomrule
\end{tabular}
\caption{The three most frequently occurring \label{mytable}$\mathcal N=2$ sector beta function combinations $b^{(1)}_\alpha=\hat{\beta}_\alpha=(\hat{\beta}_{10},\hat{\beta}_8,\hat{\beta}_{8'})$ for the $SO(10)\times{SO(8)}\times{SO(8)'}$ gauge couplings in the class of models considered in \cite{FR}. Models in the first line satisfy the decompactification condition \eqref{Condition}.}
\label{mytable}
\end{table*}

To see how such models can be constructed in practice, it is instructive to return to our example model and isolate  the `parent' $\mathcal N=2$ theory corresponding to the exact $\mathcal N=2$ subsector $H_1=G_1=h_1=g_1=0$. The one loop partition function of this $\mathcal N=2$ theory can be straightforwardly extracted from \eqref{orbifoldpartitionf} by restricting to the above subsector, and multiplying by the cardinality $|\mathbb{Z}_2^{(1)}\times\mathbb Z_2^{(3)}|=4$ of the orbifold that we are effectively trivialising
\begin{align}
	\begin{split}
	Z' = &\frac{1}{\eta^{12}\bar\eta^{24}}\,\frac{1}{2^2}\sum_{h_2,H\atop g_2,G}\frac{1}{2^3}\sum_{a,k,\rho\atop b,\ell,\sigma} \frac{1}{2^2}\sum_{H_2,H_3\atop G_2,G_3} (-1)^{a+b+HG+\Phi'} \vartheta[^a_b]^2\,\vartheta[^{a+h_2}_{b+g_2}]\,\vartheta[^{a-h_2}_{b-g_2}] \\
			&  \times  \Gamma^{(1)}_{2,2}(T,U)\ \Gamma^{(2)}_{2,2}[^{H_2}_{G_2}|^{h_2}_{g_2}](T^{(2)},U^{(2)})\ \Gamma^{(3)}_{2,2}[^{H_3}_{G_3}|^{h_2}_{g_2}](T^{(3)},U^{(3)}) \\
		& \times	\ \bar\vartheta[^k_\ell]^6 \,\bar\vartheta[^{k+h_2}_{\ell+g_2}]\,\bar\vartheta[^{k-h_2}_{\ell-g_2}]
			 \,\bar\vartheta[^{\rho}_{\sigma}]^4 \,\bar\vartheta[^{\rho+H}_{\sigma+G}]^4 \,,
	\end{split}
	\label{exactN=2part}
\end{align}
where the phase $\Phi'$ is simply obtained from $\Phi$ by setting $H_1=G_1=h_1=g_1=0$. This theory is then recognised as the ${\rm E}_8\times {\rm E}_8$ heterotic string compactified on the $T^6/(\mathbb Z_2)^4$ orbifold
\begin{equation}
	\begin{split}
			&\mathbb Z_2^{(2)} \ :\ (-1)^{F_2}\,r_2 \ ,\ r_2\,:\,\{ X^{3,4,5,6}\to-X^{3,4,5,6}\}+{\rm standard\,\, embedding}\,,\\
			&\mathbb Z_2^{(4)} \ :\  \delta_2:\,\{ X^3\to X^3+\pi R_3\}\,,\\
			&\mathbb Z_2^{(5)} \ :\ (-1)^{F_1+F_2}\,\delta_3 \ ,\ \delta_3\,:\,\{ X^5\to X^5+\pi R_5\}\,,\\
			&\mathbb Z_2^{(6)} \ :\ (-1)^{F_1}\,r_3 \ ,\ r_3\,:\,\{ \bar\phi^{5,6,7,8}\to-\bar\phi^{5,6,7,8}\} \,,\\
	\end{split}
\end{equation}
with the following choice of discrete torsion 
\begin{equation}
	\epsilon(2,4),\,\epsilon(2,6),\,\epsilon(4,6),\,\epsilon(5,6)\,.
\end{equation}
Alternatively, one could have simply considered the decompactification limit $T_2\to \infty$ of \eqref{orbifoldpartitionf} and recovered the six dimensional analogue of \eqref{exactN=2part}.

For a generic $\mathcal N=2$ model of this class, based on $\mathbb Z_2$ orbifolds, the condition for the absence of linear volume growth in gauge thresholds reads
\begin{equation}
	\beta_a^{\mathcal N=2}=12k_a\,,
	\label{CondN=2}
\end{equation}
and, hence, differs from the one given in \eqref{Condition}, which is valid for theories with $\mathcal N=1\to 0$ spontaneous supersymmetry breaking, by a factor of four. 
This multiplicative factor corresponds precisely to the cardinality of the additional orbifold group factors $\mathbb Z_2^{(1)}\times\mathbb  Z_2^{(3)}$, which are trivialised in order to obtain the parent $\mathcal N=2$ theory. For an ${\rm SO}(2n)$ gauge group at level $k=1$, the $\mathcal N=2$ beta function is straightforwardly determined by the multiplicities $n_V,n_S$ of hypermultiplets  transforming in the vectorial (fundamental) or the spinorial representation of ${\rm SO}(2n)$, respectively, so that  \eqref{CondN=2}  becomes
\begin{equation}
	n_V+2^{n-4}n_S = 2n+4 \,.
	\label{nVnScondition}
\end{equation}
In the case of ${\rm SO}(12)$ for the class of models under consideration, the untwisted sector provides at least four vectorials while $n_S\in 2\mathbb Z$, so that the only possible solutions are $(n_V,n_S)=(8,2)$ and $(n_V,n_S)=(16,0)$.

 Indeed, this is the case for the particular $\mathcal N=2$ model that we explicitly constructed here. The ${\rm E}_8\times{\rm E}_8$ gauge group is now broken to ${\rm SO}(12)\times{\rm SU}(2)\times{\rm SU}(2)'\times{\rm SO}(8)\times{\rm SO}(8)'$ with all factors realised at level $k_a=1$ and the relevant massless spectrum reads
 \begin{equation}
	\begin{array}{l l l l l}
 	 1\times ({\bf 12},{\bf 2},{\bf 2},1,1)  &,&  2\times({\bf 12}, {\bf 2},1,1,1) &,& 2\times ({\bf 32},1,1,1,1) \,, \\
	 4\times(1,{\bf 2},1,{\bf 8},1) &,&  2\times(1,1,{\bf 2},{\bf 8},1) &,& 4\times(1,{\bf 2},1,1,{\bf 8}) \,, \\
	 2\times(1,1,{\bf 2},1,{\bf 8}) &,&  2\times(1,{\bf 2},1,1,1) &,& 6\times(1,1,{\bf 2},1,1) \,.\\
	\end{array}
 \end{equation}
From the knowledge of these states, one may explicitly calculate the beta function coefficients and find
\begin{equation}
	\beta_{\rm SO(12)}^{\mathcal N=2} = 12 \quad,\quad \beta_{\rm SU(2)}^{\mathcal N=2} = 110 \quad,\quad \beta_{\rm SU(2)'}^{\mathcal N=2} = 58 \quad,\quad \beta_{\rm SO(8)}^{\mathcal N=2} = 12 \quad,\quad \beta_{\rm SO(8)'}^{\mathcal N=2} = 12  \,.
\end{equation}
Therefore, aside from the SU(2)'s which satisfy $\beta^{\mathcal N=2}>12$ and become part of the hidden group, all remaining gauge group factors  satisfy the $\mathcal N=2$ decompactification condition \eqref{CondN=2}, in accordance with the fact that the descendant $\mathcal N=0$ theory has logarithmic thresholds, as we have already seen.

This analysis suggests that a very efficient method for constructing chiral string models with well-behaved one loop gauge thresholds at large volume, is to start from an $\mathcal N=2$ theory whose spectrum satisfies \eqref{nVnScondition} for the gauge group factors from which the observable couplings will eventually arise, and then suitably orbifolding the theory to produce the desired (possibly non-supersymmetric) chiral theory of interest.


\section{Conclusions}\label{SecConclude}

The decompactification problem generically expresses that fact that, whenever the Kaluza--Klein scale $M_{\rm KK}=1/T_2$ is much smaller than the string scale, the theory is effectively six dimensional and couplings therefore have dimension of squared length. This  manifests itself in the linear growth of string thresholds with $T_2$ and, depending on the sign of the corresponding beta function, may either completely decouple  or drive the gauge group to the strong coupling regime.

In this work we proposed that a natural solution to the decompactification problem is actually already included in string theory. This is achieved if the  linear volume growth controlled by the beta function of the exact $\mathcal N=2$ subsector of the theory (or, alternatively, the six dimensional limit of the theory) is cancelled against a similar volume growth in the universal part $Y$. We  derived the conditions for this to occur, which place strong constraints on the structure of the exact  $\mathcal N=2$ subsector of the theory, and provided  explicit examples that illustrate how non-trivial solutions  with spontaneously broken (or even unbroken) supersymmetry may be constructed such that their thresholds have logarithmic, rather than polynomial,  dependence in the Scherk--Schwarz volume.

In this sense, we propose that the decompactification problem is in fact not a problem but, rather, provides a non-trivial selection criterion for realistic model building.
Whether there exist realistic string models satisfying the minimal set of desired properties outlined in the Introduction, and which do not suffer from the decompactification problem, is an interesting open question which will be investigated in future work.  Nevertheless, preliminary scans in the class of models studied in \cite{FR} indicate that some 40\% of models are in principle compatible with our proposal.



\section*{Acknowledgements}

We would like to thank C.~Angelantonj, C.~Condeescu, A.~Goudelis, C.~Kounnas and ~K. Petraki for useful discussions. I.F. wishes to thank the Theory Division of CERN and the University of Ioannina for their warm hospitality, during the final stages of this work.

\bibliographystyle{utphys}

\begin{thebibliography}{10}

\bibitem{Rohm:1983aq}
  R.~Rohm,
  ``Spontaneous Supersymmetry Breaking in Supersymmetric String Theories,''
  Nucl.\ Phys.\ B {\bf 237} (1984) 553.
  doi:10.1016/0550-3213(84)90007-5

\bibitem{Kounnas:1988ye}
  C.~Kounnas and M.~Porrati,
  ``Spontaneous Supersymmetry Breaking in String Theory,''
  Nucl.\ Phys.\ B {\bf 310} (1988) 355.
  doi:10.1016/0550-3213(88)90153-8

\bibitem{Ferrara:1988jx}
  S.~Ferrara, C.~Kounnas, M.~Porrati and F.~Zwirner,
  ``Superstrings with Spontaneously Broken Supersymmetry and their Effective Theories,''
  Nucl.\ Phys.\ B {\bf 318} (1989) 75.
  doi:10.1016/0550-3213(89)90048-5

\bibitem{Kounnas:1989dk}
  C.~Kounnas and B.~Rostand,
  ``Coordinate Dependent Compactifications and Discrete Symmetries,''
  Nucl.\ Phys.\ B {\bf 341} (1990) 641.
  doi:10.1016/0550-3213(90)90543-M

\bibitem{Scherk:1978ta}
  J.~Scherk and J.~H.~Schwarz,
  ``Spontaneous Breaking of Supersymmetry Through Dimensional Reduction,''
  Phys.\ Lett.\ B {\bf 82} (1979) 60.
  doi:10.1016/0370-2693(79)90425-8

\bibitem{Scherk:1979zr}
  J.~Scherk and J.~H.~Schwarz,
  ``How to Get Masses from Extra Dimensions,''
  Nucl.\ Phys.\ B {\bf 153} (1979) 61.
  doi:10.1016/0550-3213(79)90592-3

\bibitem{Cremmer:1983bf}
  E.~Cremmer, S.~Ferrara, C.~Kounnas and D.~V.~Nanopoulos,
  ``Naturally Vanishing Cosmological Constant in $\mathcal N$=1 Supergravity,''
  Phys.\ Lett.\ B {\bf 133} (1983) 61.
  doi:10.1016/0370-2693(83)90106-5

\bibitem{Dixon:1990pc}
  L.~J.~Dixon, V.~Kaplunovsky and J.~Louis,
  ``Moduli dependence of string loop corrections to gauge coupling constants,''
  Nucl.\ Phys.\ B {\bf 355} (1991) 649.
  doi:10.1016/0550-3213(91)90490-O
  
\bibitem{Florakis:2013ura}
  I.~Florakis,
  ``One-loop Amplitudes as BPS state sums,''
  PoS Corfu {\bf 2012} (2013) 101
  [arXiv:1303.3788 [hep-th]].

\bibitem{Angelantonj:2011br}
  C.~Angelantonj, I.~Florakis and B.~Pioline,
  ``A new look at one-loop integrals in string theory,''
  Commun.\ Num.\ Theor.\ Phys.\  {\bf 6} (2012) 159
  doi:10.4310/CNTP.2012.v6.n1.a4
  [arXiv:1110.5318 [hep-th]].

\bibitem{Angelantonj:2012gw}
  C.~Angelantonj, I.~Florakis and B.~Pioline,
  ``One-Loop BPS amplitudes as BPS-state sums,''
  JHEP {\bf 1206} (2012) 070
  doi:10.1007/JHEP06(2012)070
  [arXiv:1203.0566 [hep-th]].

\bibitem{Angelantonj:2013eja}
  C.~Angelantonj, I.~Florakis and B.~Pioline,
  ``Rankin-Selberg methods for closed strings on orbifolds,''
  JHEP {\bf 1307} (2013) 181
  doi:10.1007/JHEP07(2013)181
  [arXiv:1304.4271 [hep-th]].

\bibitem{Angelantonj:2015rxa}
  C.~Angelantonj, I.~Florakis and B.~Pioline,
  ``Threshold corrections, generalised prepotentials and Eichler integrals,''
  Nucl.\ Phys.\ B {\bf 897} (2015) 781
  doi:10.1016/j.nuclphysb.2015.06.009
  [arXiv:1502.00007 [hep-th]].

\bibitem{Florakis:2016boz}
  I.~Florakis and B.~Pioline,
  ``On the Rankin-Selberg method for higher genus string amplitudes,''
  arXiv:1602.00308 [hep-th].

\bibitem{Antoniadis:1990ew}
  I.~Antoniadis,
  ``A Possible new dimension at a few TeV,''
  Phys.\ Lett.\ B {\bf 246} (1990) 377.
  doi:10.1016/0370-2693(90)90617-F

\bibitem{Abel:2016hgy}
  S.~Abel,
  ``A dynamical mechanism for large volumes with consistent couplings,''
  JHEP {\bf 1611} (2016) 085
  doi:10.1007/JHEP11(2016)085
  [arXiv:1609.01311 [hep-th]].

\bibitem{Abel:2017rch}
  S.~Abel and R.~J.~Stewart,
  ``On exponential suppression of the cosmological constant in non-SUSY strings at two loops and beyond,''
  arXiv:1701.06629 [hep-th].

\bibitem{Ginsparg:1986wr}
  P.~H.~Ginsparg and C.~Vafa,
  ``Toroidal Compactification of Nonsupersymmetric Heterotic Strings,''
  Nucl.\ Phys.\ B {\bf 289} (1987) 414.
  doi:10.1016/0550-3213(87)90387-7

\bibitem{Itoyama:1986ei}
  H.~Itoyama and T.~R.~Taylor,
  ``Supersymmetry Restoration in the Compactified O(16) $\times$ O(16)-prime Heterotic String Theory,''
  Phys.\ Lett.\ B {\bf 186} (1987) 129.
  doi:10.1016/0370-2693(87)90267-X

\bibitem{Angelantonj:2006ut}
  C.~Angelantonj, M.~Cardella and N.~Irges,
  ``An Alternative for Moduli Stabilisation,''
  Phys.\ Lett.\ B {\bf 641} (2006) 474
  doi:10.1016/j.physletb.2006.08.072
  [hep-th/0608022].

\bibitem{Fischler:1986ci}
  W.~Fischler and L.~Susskind,
  ``Dilaton Tadpoles, String Condensates and Scale Invariance,''
  Phys.\ Lett.\ B {\bf 171} (1986) 383.
  doi:10.1016/0370-2693(86)91425-5
  
\bibitem{Fischler:1986tb}
  W.~Fischler and L.~Susskind,
  ``Dilaton Tadpoles, String Condensates and Scale Invariance. 2.,''
  Phys.\ Lett.\ B {\bf 173} (1986) 262.
  doi:10.1016/0370-2693(86)90514-9

\bibitem{McClain:1986id}
  B.~McClain and B.~D.~B.~Roth,
  ``Modular Invariance for Interacting Bosonic Strings at Finite Temperature,''
  Commun.\ Math.\ Phys.\  {\bf 111} (1987) 539.
  doi:10.1007/BF01219073
  
\bibitem{O'Brien:1987pn}
  K.~H.~O'Brien and C.~I.~Tan,
  ``Modular Invariance of Thermopartition Function and Global Phase Structure of Heterotic String,''
  Phys.\ Rev.\ D {\bf 36} (1987) 1184.
  doi:10.1103/PhysRevD.36.1184

\bibitem{Atick:1988si}
  J.~J.~Atick and E.~Witten,
  ``The Hagedorn Transition and the Number of Degrees of Freedom of String Theory,''
  Nucl.\ Phys.\ B {\bf 310} (1988) 291.
  doi:10.1016/0550-3213(88)90151-4

\bibitem{Kutasov:1990sv}
  D.~Kutasov and N.~Seiberg,
  ``Number of degrees of freedom, density of states and tachyons in string theory and CFT,''
  Nucl.\ Phys.\ B {\bf 358} (1991) 600.
  doi:10.1016/0550-3213(91)90426-X

\bibitem{Antoniadis:1991kh}
  I.~Antoniadis and C.~Kounnas,
  ``Superstring phase transition at high temperature,''
  Phys.\ Lett.\ B {\bf 261} (1991) 369.
  doi:10.1016/0370-2693(91)90442-S

\bibitem{Antoniadis:1999gz}
  I.~Antoniadis, J.~P.~Derendinger and C.~Kounnas,
  ``Nonperturbative temperature instabilities in $\mathcal N=4$ strings,''
  Nucl.\ Phys.\ B {\bf 551} (1999) 41
  doi:10.1016/S0550-3213(99)00171-6
  [hep-th/9902032].

\bibitem{Dienes:1994np}
  K.~R.~Dienes,
  ``Modular invariance, finiteness, and misaligned supersymmetry: New constraints on the numbers of physical string states,''
  Nucl.\ Phys.\ B {\bf 429} (1994) 533
  doi:10.1016/0550-3213(94)90153-8
  [hep-th/9402006].

\bibitem{Angelantonj:2008fz}
  C.~Angelantonj, C.~Kounnas, H.~Partouche and N.~Toumbas,
  ``Resolution of Hagedorn singularity in superstrings with gravito-magnetic fluxes,''
  Nucl.\ Phys.\ B {\bf 809} (2009) 291
  doi:10.1016/j.nuclphysb.2008.10.010
  [arXiv:0808.1357 [hep-th]].

\bibitem{Angelantonj:2010ic}
  C.~Angelantonj, M.~Cardella, S.~Elitzur and E.~Rabinovici,
  ``Vacuum stability, string density of states and the Riemann zeta function,''
  JHEP {\bf 1102} (2011) 024
  doi:10.1007/JHEP02(2011)024
  [arXiv:1012.5091 [hep-th]].

\bibitem{Florakis:2010ty}
  I.~Florakis, C.~Kounnas and N.~Toumbas,
  ``Marginal Deformations of Vacua with Massive boson-fermion Degeneracy Symmetry,''
  Nucl.\ Phys.\ B {\bf 834} (2010) 273
  doi:10.1016/j.nuclphysb.2010.03.020
  [arXiv:1002.2427 [hep-th]].

\bibitem{FR}
  I.~Florakis and J.~Rizos,
  ``Chiral Heterotic Strings with Positive Cosmological Constant,''
  Nucl.\ Phys.\ B {\bf 913} (2016) 495
  doi:10.1016/j.nuclphysb.2016.09.018
  [arXiv:1608.04582 [hep-th]].

\bibitem{Abel:2015oxa}
  S.~Abel, K.~R.~Dienes and E.~Mavroudi,
  ``Towards a nonsupersymmetric string phenomenology,''
  Phys.\ Rev.\ D {\bf 91} (2015) no.12,  126014
  doi:10.1103/PhysRevD.91.126014
  [arXiv:1502.03087 [hep-th]].

\bibitem{Aaronson:2016kjm}
  B.~Aaronson, S.~Abel and E.~Mavroudi,
  ``On interpolations from SUSY to non-SUSY strings and their properties,''
  arXiv:1612.05742 [hep-th].

\bibitem{Kounnas:2016gmz}
  C.~Kounnas and H.~Partouche,
  ``Super no-scale models in string theory,''
  Nucl.\ Phys.\ B {\bf 913} (2016) 593
  doi:10.1016/j.nuclphysb.2016.10.001
  [arXiv:1607.01767 [hep-th]].

\bibitem{Kounnas:2017mad}
  C.~Kounnas and H.~Partouche,
  ``$\mathcal N=2 \to 0$ super no-scale models and moduli quantum stability,''
  arXiv:1701.00545 [hep-th].

\bibitem{Kaplunovsky:1987rp}
  V.~S.~Kaplunovsky,
  ``One Loop Threshold Effects in String Unification,''
  Nucl.\ Phys.\ B {\bf 307} (1988) 145
   Erratum: [Nucl.\ Phys.\ B {\bf 382} (1992) 436]
  doi:10.1016/0550-3213(92)90193-F, 10.1016/0550-3213(88)90526-3
  [hep-th/9205068].

\bibitem{Angelantonj:2014dia}
  C.~Angelantonj, I.~Florakis and M.~Tsulaia,
  ``Universality of Gauge Thresholds in Non-Supersymmetric Heterotic Vacua,''
  Phys.\ Lett.\ B {\bf 736} (2014) 365
  doi:10.1016/j.physletb.2014.08.001
  [arXiv:1407.8023 [hep-th]].

\bibitem{Florakis:2015txa}
  I.~Florakis,
  ``Universality of radiative corrections to gauge couplings for strings with spontaneously broken supersymmetry,''
  J.\ Phys.\ Conf.\ Ser.\  {\bf 631} (2015) no.1,  012079
  doi:10.1088/1742-6596/631/1/012079
  [arXiv:1502.07537 [hep-th]].

\bibitem{Angelantonj:2015nfa}
  C.~Angelantonj, I.~Florakis and M.~Tsulaia,
  ``Generalised universality of gauge thresholds in heterotic vacua with and without supersymmetry,''
  Nucl.\ Phys.\ B {\bf 900} (2015) 170
  doi:10.1016/j.nuclphysb.2015.09.007
  [arXiv:1509.00027 [hep-th]].

\bibitem{Angelantonj:2016ibb}
  C.~Angelantonj, I.~Florakis and M.~Tsulaia,
  ``Universality in radiative corrections for non-supersymmetric heterotic vacua,''
  PoS PLANCK {\bf 2015} (2016) 005.

\bibitem{Florakis:2016aoi}
  I.~Florakis,
  ``Gravitational Threshold Corrections in Non-Supersymmetric Heterotic Strings,''
  Nucl.\ Phys.\ B {\bf 916} (2017) 484
  doi:10.1016/j.nuclphysb.2017.01.016
  [arXiv:1611.10323 [hep-th]].

\bibitem{Kiritsis:1996xd}
  E.~Kiritsis, C.~Kounnas, P.~M.~Petropoulos and J.~Rizos,
  ``Solving the decompactification problem in string theory,''
  Phys.\ Lett.\ B {\bf 385} (1996) 87
  doi:10.1016/0370-2693(96)00880-5
  [hep-th/9606087].

\bibitem{Faraggi:2014eoa}
  A.~E.~Faraggi, C.~Kounnas and H.~Partouche,
  ``Large volume susy breaking with a solution to the decompactification problem,''
  Nucl.\ Phys.\ B {\bf 899} (2015) 328
  doi:10.1016/j.nuclphysb.2015.08.001
  [arXiv:1410.6147 [hep-th]].
  
\bibitem{Harvey:1995fq}
  J.~A.~Harvey and G.~W.~Moore,
  ``Algebras, BPS states, and strings,''
  Nucl.\ Phys.\ B {\bf 463} (1996) 315
  doi:10.1016/0550-3213(95)00605-2
  [hep-th/9510182].

\bibitem{Harvey:1996gc}
  J.~A.~Harvey and G.~W.~Moore,
  ``On the algebras of BPS states,''
  Commun.\ Math.\ Phys.\  {\bf 197} (1998) 489
  doi:10.1007/s002200050461
  [hep-th/9609017].

\bibitem{Kiritsis:1996dn}
  E.~Kiritsis, C.~Kounnas, P.~M.~Petropoulos and J.~Rizos,
  ``Universality properties of N=2 and N=1 heterotic threshold corrections,''
  Nucl.\ Phys.\ B {\bf 483} (1997) 141
  doi:10.1016/S0550-3213(96)00550-0
  [hep-th/9608034].
  
\bibitem{Antoniadis:1992sa}
  I.~Antoniadis, E.~Gava and K.~S.~Narain,
  ``Moduli corrections to gravitational couplings from string loops,''
  Phys.\ Lett.\ B {\bf 283} (1992) 209
  doi:10.1016/0370-2693(92)90009-S
  [hep-th/9203071].

\bibitem{Antoniadis:1992rq}
  I.~Antoniadis, E.~Gava and K.~S.~Narain,
  ``Moduli corrections to gauge and gravitational couplings in four-dimensional superstrings,''
  Nucl.\ Phys.\ B {\bf 383} (1992) 93
  doi:10.1016/0550-3213(92)90672-X
  [hep-th/9204030].

\bibitem{Kiritsis:1994ta}
  E.~Kiritsis and C.~Kounnas,
  ``Infrared regularization of superstring theory and the one loop calculation of coupling constants,''
  Nucl.\ Phys.\ B {\bf 442} (1995) 472
  doi:10.1016/0550-3213(95)00156-M
  [hep-th/9501020].

\bibitem{Petropoulos:1996rr}
  P.~M.~Petropoulos and J.~Rizos,
  ``Universal moduli dependent string thresholds in Z(2) x Z(2) orbifolds,''
  Phys.\ Lett.\ B {\bf 374} (1996) 49
  doi:10.1016/0370-2693(96)00230-4
  [hep-th/9601037].
  
\bibitem{Henningson:1996jz}
    M.~Henningson and G.~W.~Moore,
    ``Threshold corrections in K3 x T2 heterotic string compactifications,''
    Nucl.\ Phys.\ B {\bf 482} (1996) 187
    doi:10.1016/S0550-3213(96)00549-4
    [hep-th/9608145].

\bibitem{Kiritsis:1997hf}
  E.~Kiritsis and N.~A.~Obers,
  ``Heterotic type I duality in D $<$ 10-dimensions, threshold corrections and D instantons,''
  JHEP {\bf 9710} (1997) 004
  doi:10.1088/1126-6708/1997/10/004
  [hep-th/9709058].
  




  
  
  
  

\end{thebibliography}
\providecommand{\href}[2]{#2}\begingroup\raggedright\endgroup

\end{document}